\documentclass[iop, revtex4, twocolumn, appendixfloats, numberedappendix]{emulateapj}

\usepackage{graphicx}
\usepackage{subfigure}

\usepackage{hyperref}
\usepackage{natbib}
\usepackage{amsmath}
\usepackage{mathrsfs}

\begin{document}
\title{New Constraints on Early-Type Galaxy Assembly from Spectroscopic 
Metallicities of Globular Clusters in M87}

\author{Alexa Villaume}
\affiliation{Department of Astronomy \& Astrophysics, 
		University of California Santa Cruz,
		1156 High Street, Santa Cruz, CA 95064, USA} 

\author{Aaron J. Romanowsky} 
\affiliation{Department of Physics \& Astronomy, 
		San Jose State University, 
		One Washington Square, San Jose, CA 95192, USA}
\affiliation{University of California Observatories, 
		1156 High Street, Santa Cruz, CA, 95064, CA, USA}

\author{Jean Brodie}
\affiliation{Department of Astronomy \& Astrophysics, 
		University of California Santa Cruz,
		1156 High Street, Santa Cruz, CA 95064, USA} 
		
\author{Jay Strader}
\affiliation{Center for Data Intensive and Time Domain Astronomy,
		Department of Physics and Astronomy,
		Michigan State University, 
		East Lansing, MI 48824, USA}

\begin{abstract}

The observed characteristics of globular cluster (GC) systems, such as metallicity distributions, are commonly used to place constraints on galaxy formation models. However, obtaining reliable metallicity values is particularly difficult because of our limited means to obtain high quality spectroscopy of extragalactic GCs. Often, ``color--metallicity relations'' are  invoked to convert easier-to-obtain photometric measurements into metallicities, but there is no consensus on what form these relations should take. In this paper we make use of multiple photometric datasets and iron metallicity values derived from applying full-spectrum stellar population synthesis models to deep Keck/LRIS spectra of 177 GCs centrally located around M87 to obtain a new color--metallicity relation. Our new relation differs substantially from previous relations in the blue, and we present evidence that the M87 relation differs from that of the Milky Way GCs, suggesting environmental dependence of GC properties. We use our color--metallicity relation to derive a new GC metallicity-host galaxy luminosity relation for red and blue GCs and find a shallower relation for the blue GCs than what previous work has found and that the metal-poor GCs are more enriched than what was previously found. This could indicate that the progenitor satellite galaxies that now make up the stellar halos of early-type galaxies are more massive and formed later than previously thought, or that the properties of metal-poor GCs are less dependent on their present-day host, indicating a common origin.
\end{abstract}

\section{Introduction}

Although $\Lambda$CDM cosmology gives us the broad framework that galaxies form hierarchically, the details of how giant early-type galaxies (ETGs) form is still a matter of debate. Areas of ongoing uncertainty include the {\it assembly} of ETGs such as the epoch of the last merger and what kind of progenitor galaxies now constitute the stellar halos of ETGs. In particular, while cosmological simulations point to massive progenitor satellites as building the stellar halos of present day giant ETGs \citep[see, for example, Figure 13 of][]{pillepich2018}, observational constraints suggest dwarf galaxies as the progenitors \citep[Figure 2 of][]{forbes2015}. 

Globular clusters (GCs) are nearly ubiquitous around galaxies and have been determined to be old ($\sim 10$ Gyr) in a variety of systems \citep[see references in][]{brodie2006}. Those properties as well as their luminosity ($-5 < M_V < -10$) make them potentially useful tracers of galaxy formation and assembly. However, the promise of GCs in this capacity has yet to be fully realized, in part, because of our limited means to understand the present-day physical properties of GC systems. 

\citet{bergh1975} first used the likely connection between a galaxy's star-formation episodes and its GC population to suggest a link between galaxy luminosity and the metallicities of its GCs. This relation was confirmed by \citet{brodie1991}, and subsequently the paradigm of {\it bimodality} has overtaken the extragalactic GC field. Bimodality was first established through optical color distributions from {\it Hubble Space Telescope} photometry \citep{gebhardt1999, kundu2001a, larsen2001}. Since then GC systems around ETGs are treated as composed from two subpopulations and separately track the subpopulation characteristics with host galaxy characteristics to place constraints on galaxy formation scenarios \citep[e.g.,][]{cote2002, strader2005, rhode2005, li2014}. Recently though, \citet{harris2017a} presented observational evidence that the most massive ETGs, brightest cluster galaxies, can have broad unimodal distributions in addition to bimodal distributions.

GCs are thought to contain coeval stars with old ages and mostly homogenous metallicities and so broadband colors of GCs are generally considered to reflect their underlying mean metallicity. The simplicity of this logic belies the fact that there is no consensus on how broadband colors should be transformed into metallicities (parameterized as the ``color--metallicity relation'').  The core of almost all astronomical problems is translating observed characteristics into physically meaningful properties and understanding GC systems is no exception.  We have very limited means to obtain spectroscopy -- our best observational tool for deriving physical stellar population characteristics -- of individual GCs around the largest elliptical galaxies. This is a result of a two-fold problem: at the distances of elliptical galaxies, GCs are faint, and the largest elliptical galaxies can host systems of tens of thousands of GCs. This means that in extragalactic work we often only have access to coarse observational characteristics of individual GCs, such as broadband photometry. 

The problems associated with obtaining the metallicity distribution are illustrated through the difference between the \citet{harris2006} and \citet{peng2006} color--metallicity relations. \citet{peng2006} used {\it HST}/ACS photometry of GCs around Virgo Cluster galaxies from \citet{jordan2004} and metallicities gathered from the few spectroscopic studies of extragalactic GCs available at the time \citep[][]{cohen1998, cohen2003}. \citet{peng2006} found a color--metallicity relation with a significant break when transitioning to the blue GCs, but, crucially their relation was based almost entirely on Milky Way GCs at the metal-poor end. \citet{harris2006} derived a linear relation between $B-I$ colors and metallicities for Milky Way GCs to interpret the broadband colors they obtained for Virgo Cluster GC systems. \citet{peng2006} and \citet{harris2006} reported essentially the same color distributions for the Virgo GC systems but different {\it metallicity} distributions.

Despite their differences, both \citet{peng2006} and \citet{harris2006} maintained evidence for metallicity bimodality but that paradigm was challenged by \citet{yoon2006}. \citet{yoon2006} introduced the idea of generating synthetic color--metallicity relations to transform the overall color distributions of GC systems to metallicity distributions. They found that highly non-linear color--metallicity relations, like those that result from inclusion of helium-rich hot horizontal branch stars, can transform unimodal metallicity distributions into bimodal color distributions.

Contrary to \citet{yoon2006} and their follow-up work \citep{lee2019}, studies that directly model the spectroscopic observations of GCs consistently find bimodal metallicity distributions \citep{alves-brito2011, usher2012, brodie2012}. Despite the near-consensus regarding bimodality, the differences in various color--metallicity relations \citep[see also][]{usher2012} highlight that there may be physical properties beyond metallicity that affect the broadband colors of GCs.

%But the differences between the \citet{harris2006} and \citet{peng2006} color--metallicity relations and the color--metallicity relations determined for the systems included in the \citet{usher2012} sample highlight that there appear to be other physical properties that affect broadband colors than just metallicity.

Full-spectrum stellar population synthesis (SPS) modeling provides a way to move past these problems. Modern full-spectrum models allow for variations in abundance patterns \citep[][]{conroy2014} over a variety of ages \citep[][]{choi2014} and metallicities \citep[][]{conroy2018a}. In addition to fully accounting for possible variations in many stellar population parameters, we have shown that full-spectrum fitting allows us to extract information from data in a lower signal-to-noise (S/N) regime than traditional index fitting \citep[][]{conroy2018a}.

It is exactly this last property of full-spectrum SPS models that enables us to make use of the \citet{strader2011} database of spectroscopy of individual GCs around M87. In this paper we present the most comprehensive and accurate compendium of metallicities for individual GCs around M87 (which we describe in Section~\ref{sec:model}). We use these metallicities to derive a new color--metallicity relation in Section~\ref{sec:results}. We discuss the implications of the new color--metallicity relation in Section~\ref{sec:discussion}.

\begin{figure*}
\includegraphics[width=1\textwidth]{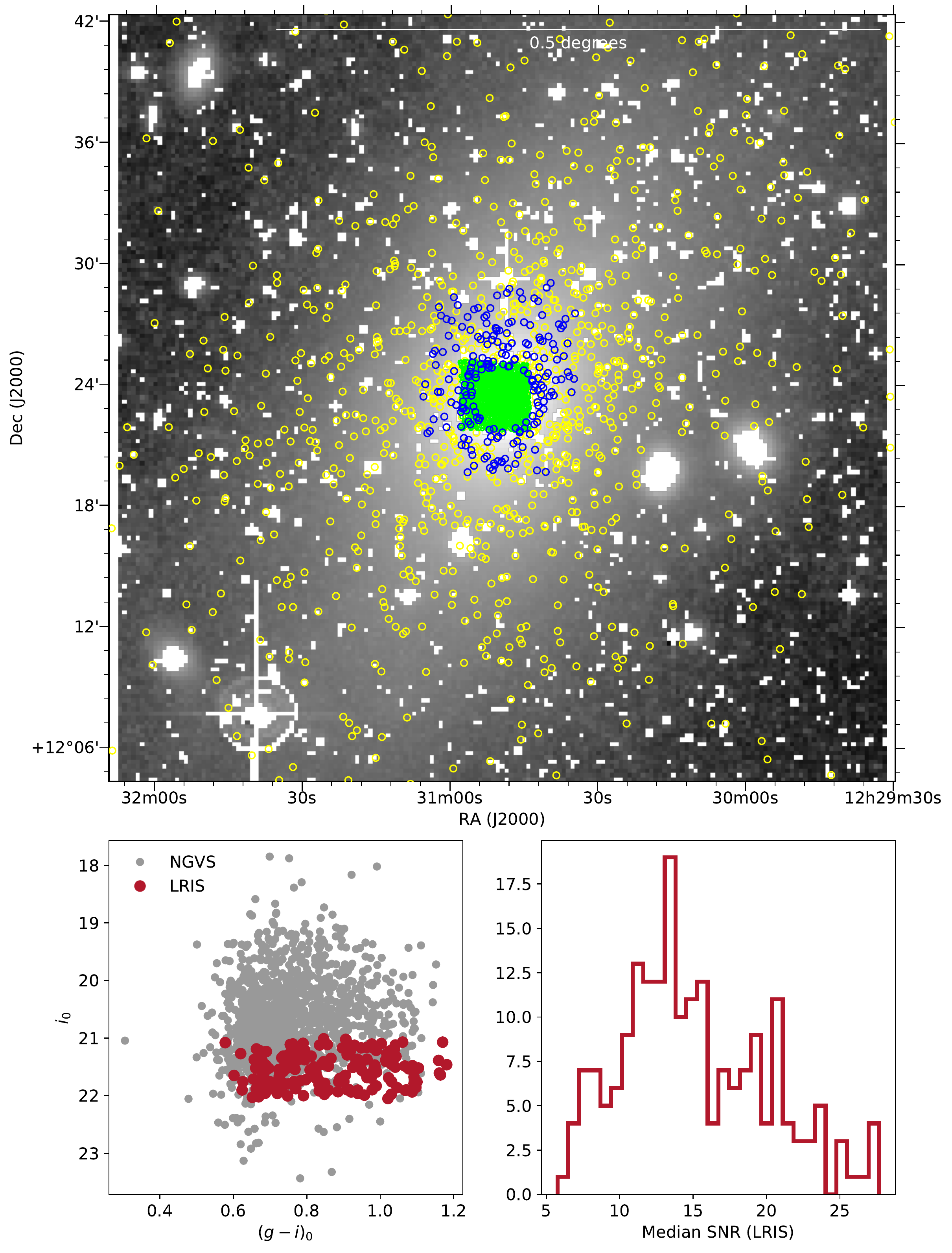}
\caption{(Upper) Image of M87 from the Burrell Schmidt Deep Virgo Survey \citep{mihos2017} with spatial distributions of the NGVS photometry (yellow), ACSVCS photometry (green), and LRIS spectroscopy (blue), %Hectospec data (red) and Virus-P data () marked.
 (Left) Color--magnitude diagram for the M87 GCs from the NGVS catalog from \citet{oldham2016} (grey), the culled sample of the LRIS data set (red) from \citet{strader2011}. (Right) Histogram of median signal-to-noise ratio values for the individual spectra of the LRIS sample.}
\label{fig:catalogs}
\end{figure*}

\section{Stellar Population Synthesis Modeling}
\label{sec:model}

We make use of the Keck/LRIS spectroscopic subsample of the dataset described in \citet{strader2011} ($\sim 3300 - 5600 {\rm \AA}$). In the top panel of Figure~\ref{fig:catalogs} we show a deep image of M87 from the Burrell Schmidt Deep Virgo Survey \citep{mihos2017} with the NGVS photometric catalog \citep[yellow,][]{oldham2016}, the ACSVCS photometric catalog \citep[green,][]{jordan2009}, and the LRIS spectroscopic sample \citep[blue,][]{strader2011}.

\begin{figure*}
\includegraphics[width=1.0\textwidth]{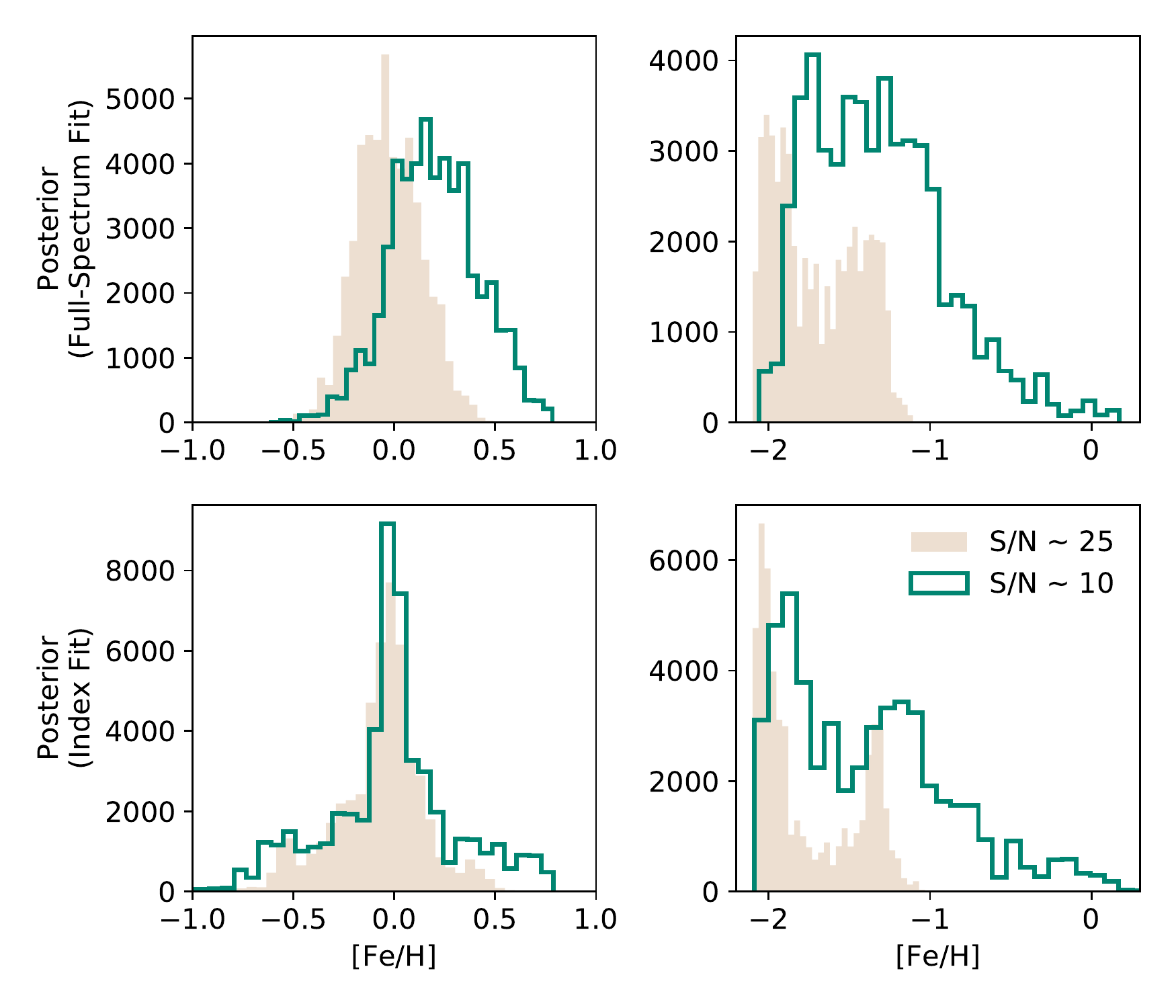}
\caption{Comparing [Fe/H] posteriors for metal-rich GCs (left column) and metal-poor GCs (right column) where the spectra were fitted using the full spectrum (top row) and Lick indices (row). In all cases the full-spectrum fits provide better constraints on [Fe/H] than index fits. For the metal-rich GCs, the index fits have broader tails than the full-spectrum fits of the same GCs. For the metal-poor GCs, the index fits do not result in well-behaved posterior distributions on [Fe/H]. The GCs in this figure correspond to the GCs in Figures~\ref{fig:model_comparison_mr} and \ref{fig:model_comparison_mp}.}
\label{fig:feh_post}
\end{figure*}

There are several features of the \citet{strader2011} sample that are salient to the work presented in this paper. First, the clusters in this sample were chosen to be fainter in magnitude than the obvious ``blue tilt'' clusters, which will help when we assess bimodality. Second, in the bottom-left panel of Figure~\ref{fig:catalogs} we compare the NGVS photometry sample with the LRIS sample in color--magnitude space. The LRIS sample spans nearly the whole color range of the M87 GC system (middle panel Figure~\ref{fig:catalogs}). 

\begin{figure*}
\includegraphics[width=1\textwidth]{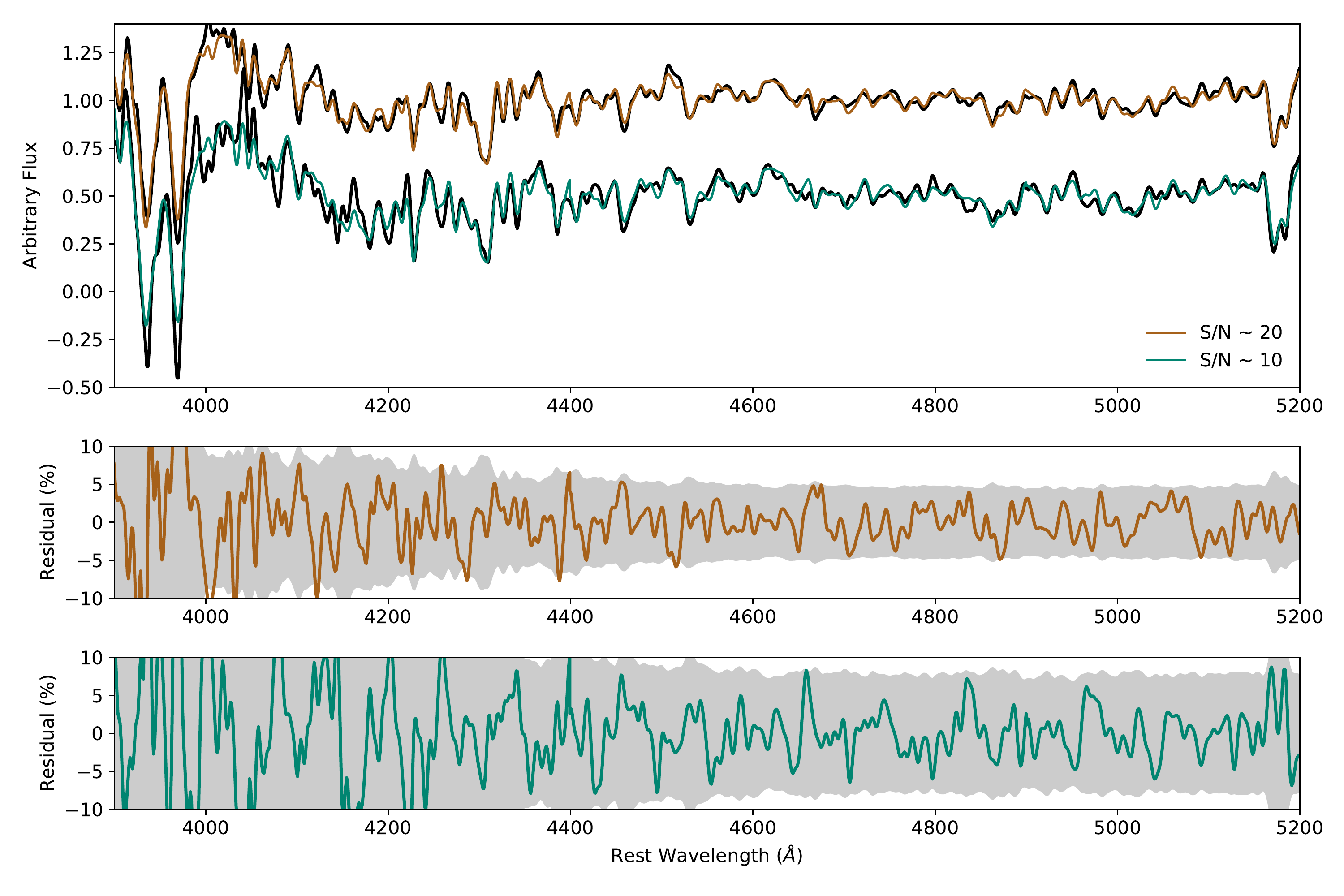}
\caption{Top: Comparison of metal-rich ([Fe/H] $> +0.1$) LRIS spectra (black) and best-fit models for a high-S/N (H51142, ($g-z$)$_{\rm NGVS} = +1.38$, brown) observation and a low-S/N (H51943, ($g-z$)$_{\rm NGVS} = +1.33$, green). Middle: Comparison of residuals between best-fit model and data for H51075 and uncertainty of flux from the input spectrum (grey). Bottom: Same as middle panel but for H51943.}
\label{fig:model_comparison_mr}
\end{figure*}

\begin{figure*}
\includegraphics[width=1\textwidth]{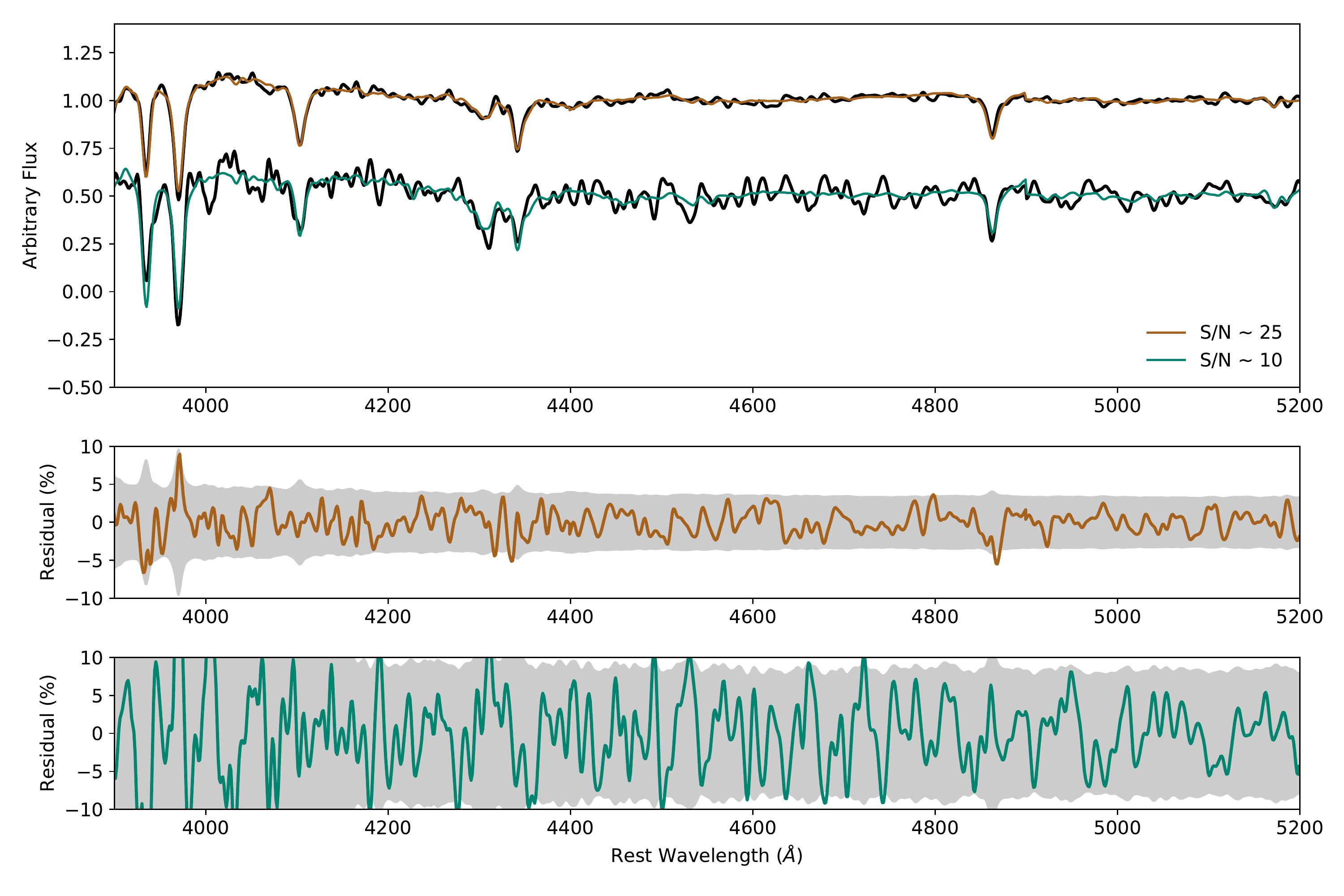}
\caption{Same as for Figure~\ref{fig:model_comparison_mr} but for the metal-poor GCs ($< -1.5$) H38032 (($g-z$)$_{\rm NGVS} = +0.70$, brown) and H42981 (($g-z$)$_{\rm NGVS} = +0.69$, green).}
\label{fig:model_comparison_mp}
\end{figure*}

This work makes use of the updated full-spectrum SPS models (\texttt{alf}) described in \citet{conroy2018a}. The most relevant update of the \citet{conroy2018a} models with regards to this work is the expansion of stellar parameter coverage of the models with the Spectral Polynomial Interpolator \citep[SPI,][]{villaume2017a}\footnote{\url{https://github.com/AlexaVillaume/SPI_Utils}}. With SPI we used the optical MILES stellar library \citep{sb2006a}, the Extended IRTF stellar library \citep[E-IRTF,][]{villaume2017a}, and a large sample of M Dwarf spectra \citep{mann2015} to create a data-driven model which we can use to generate stellar spectra as a function of effective temperature, surface gravity, and metallicity.

The empirical parameter space is set by the E-IRTF and \citet{mann2015} samples which together span $-2.0 \lesssim {\rm [Fe/H]} \lesssim +0.5$ and $3.9 \lesssim {\rm log~T_{eff}} \lesssim 3.5$. To preserve the quality of interpolation at the edges of empirical parameter space we augment the training set with a theoretical stellar library (C3K). The \texttt{alf} models allow for variable abundance patterns by differentially including theoretical element response functions. In \citet{conroy2018a} we fitted the \citet{schiavon2005} spectroscopic sample of Milky Way GCs and compared the \texttt{alf}-inferred [Fe/H] values with a compilation of [Fe/H] values from the literature \citep[see][for details]{roediger2014}. Over a range of $ -2.5 \lesssim {\rm [Fe/H]} \lesssim +0.0$ we had nearly one-to-one consistency between the literature values and our measured [Fe/H] values from integrated light (specifically, ${\rm [Fe/H]_{lit}}\propto 1.06 {\rm [Fe/H]_{\texttt{alf}} }$).

The LRIS sample is in the low signal-to-noise (S/N) regime with $\sim 5-30$ encompassing the range of the median S/N over each spectrum (bottom-right panel in Figure~\ref{fig:catalogs}). In this modest S/N regime it is difficult to obtain accurate stellar population parameters \citep{sb2011a}. To obtain an accurate color--metallicity relation we need the metallicities of {\it individual} GCs and therefore stacking spectra is not a good option for this particular problem. 

We fit  objects using both full-spectrum (left) and traditional line-index methods (right). For our line-index fits we use the canonical set of Lick indices \citep{faber1985, burstein1986, worthey1994}: H$\delta_F$, CN$_2$, Ca4227, G4300, H$\gamma_F$, Fe4383, Fe4531, C$_2$4668, H$\beta$, Fe5015, Mg$b$, Fe5270, Fe345, and Fe5406. For the full-spectrum fits we fit in simple-mode over the wavelength regions: $3900-4400{\rm \AA}$, $4400-4900{\rm \AA}$, $4900-5200{\rm \AA}$. We smoothed the LRIS spectra to be a constant 200 km/s over the whole wavelength range. 

In Figure~\ref{fig:feh_post} we demonstrate the utility of full-spectrum fitting over line-index methods. In this figure we compare [Fe/H] posteriors for metal-rich GCs (left column) and metal-poor GCs (right column) where the spectrum were fitted using the full spectrum (top row) and Lick indices (bottom row). In each panel we compare the results of high-S/N and low-S/N spectra. We demonstrate that in both the metal-rich and metal-poor cases the posteriors are better constrained when full-spectrum fitting is used. In the metal-rich case, the posterior distributions for high and low-S/N using Lick indices have larger tails than the posterior distributions from full-spectrum fitting. The real utility of the new models is shown in the low-metallicity case where the posterior distributions are more centered on a single value from full-spectrum fitting than from indices.

In Figures~\ref{fig:model_comparison_mr} and \ref{fig:model_comparison_mp} we examine the quality of our fits for metal-rich and metal-poor GCs, respectively. In each Figure we compare the LRIS spectrum (black) with the best-fit model spectrum for a high-S/N (brown) spectrum and a low-S/N (green) spectrum in the top panel. The middle and bottom panels in each figure compare the residuals between the high-S/N spectra and low-S/N, respectively, with the flux uncertainty of each LRIS spectrum (grey band). These comparisons demonstrate that the fitting was successful as the residuals are consistent with the flux uncertainty. Even with the low-S/N spectra several spectral features are still prominent, including CaII, H$\delta$, H$\beta$, and Mg$b$, which are well-characterized by the best-fit model.

After we fit every spectrum we visually inspected the residuals between the observed spectrum and the best-fit model. From this inspection we identified cases where the best-fit model is clearly a poor fit to the data. We removed these clusters from our subsequent analysis, bringing our final sample to 177 GCs. Of the 23 GCs we culled from our final metallicity sample, 20 have NGVS photometry, and 15 of those are considered to be blue ($g-z < 1.0$). This suggests that it is more difficult to obtain adequate spectra of the blue and, presumably, metal-poor GCs. However, with our remaining blue GCs we are still adequately covering the metal-poor parameter space. The posteriors for the [Fe/H] values for the final sample of GCs are available at \url{https://github.com/AlexaVillaume/m87-gc-feh-posteriors}.

\section{Results}
\label{sec:results}
\subsection{Comparison to Previous Work}

\citet{cohen1998} previously did stellar population analysis on a spectroscopic sample of M87 GCs \citep{cohen1997} using indices to determine metallicity values. To aid our analysis we matched the \citet{cohen1998} sample to the \citet{oldham2016} NGVS-based photometry catalog. We matched the \citet{cohen1998} sample to the data presented in \citet{hanes2001}, which provided right ascension and declination values for all the GCs in the \citet{strom1981} catalog that \citet{cohen1997} selected their targets from.

Then we used the position values to match with the \citet{oldham2016} catalog with a max separation of $1''$. We dereddened the \citet{oldham2016} photometry using the \citet{fitzpatrick1999} extinction law and extinction values taken from the \citet{schlegel1998} dust map using the NASA/IPAC Infrared Science Archive (${ A}_g = 0.087$,  ${ A}_i = 0.048$, ${ A}_z = 0.034$, ${ R}_g = 3.793$,  ${ R}_i = 2.086$, ${ R}_z = 1.479$).

\begin{figure}
\includegraphics[width=0.5\textwidth]{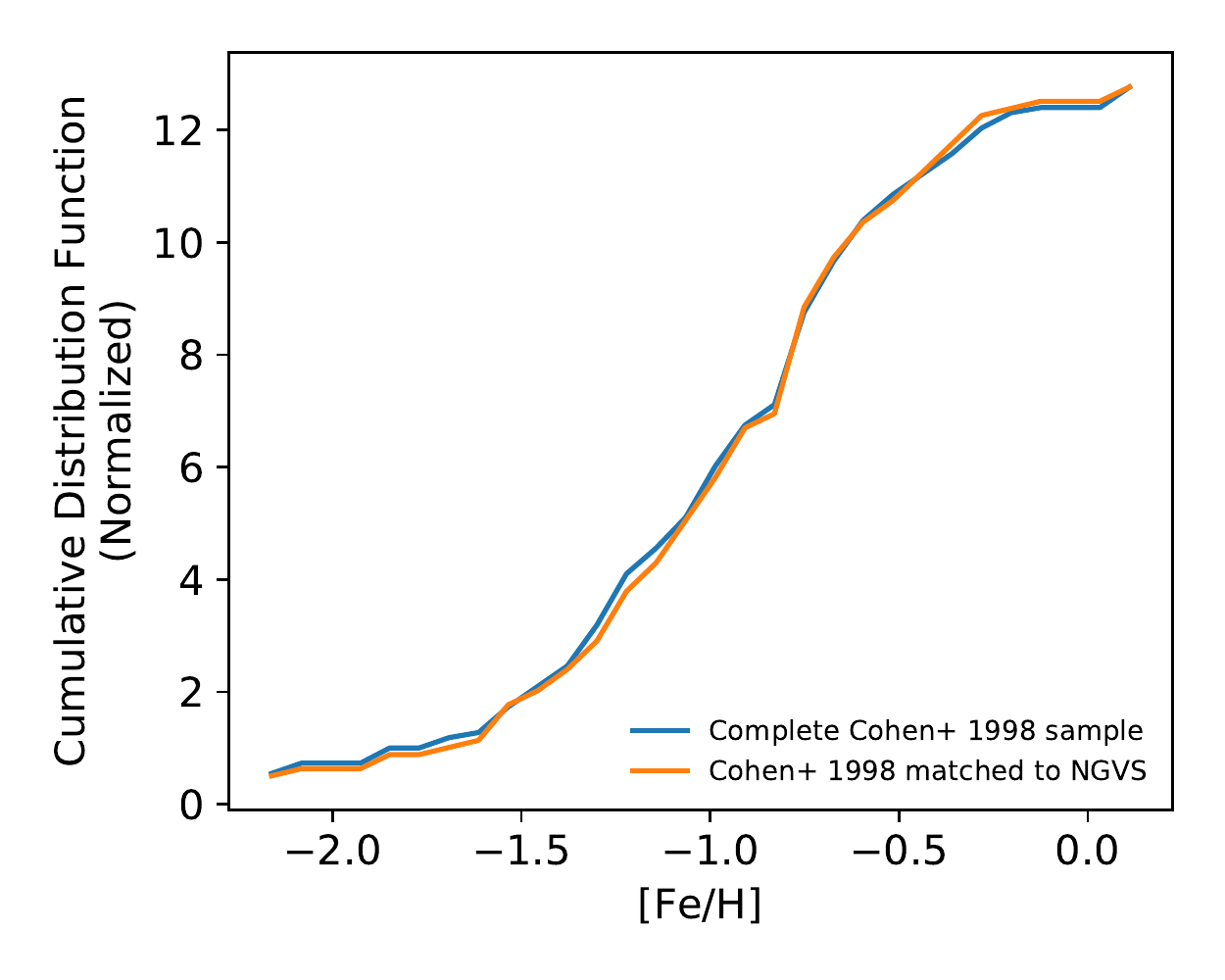}
\caption{Comparing the complete sample of M87 GC metallicities from \citet{cohen1998} (140 GCs, blue) with the sample when matched to the \citet{oldham2016} photometry (101 GCs, orange) to demonstrate that we are not biasing the \citet{cohen1998} metallicity distribution by matching to photometry.}
\label{fig:cohen_check}
\end{figure}

We do not include the objects in Table 1 of \citet{cohen1997} and not every GC in the \citet{cohen1998} sample has NGVS photometry so we go from the full \citet{cohen1998} sample of 150 GCs with [Fe/H] values  to 101 GCs. In the left panel of Figure~\ref{fig:cohen_check} we compare the normalized cumulative metallicity distributions of both the full  (blue) and matched (orange) \citet{cohen1998} sample. This comparison demonstrates that we are not biasing the \citet{cohen1998} sample by doing the matching.

\begin{figure*}
\includegraphics[width=1\textwidth]{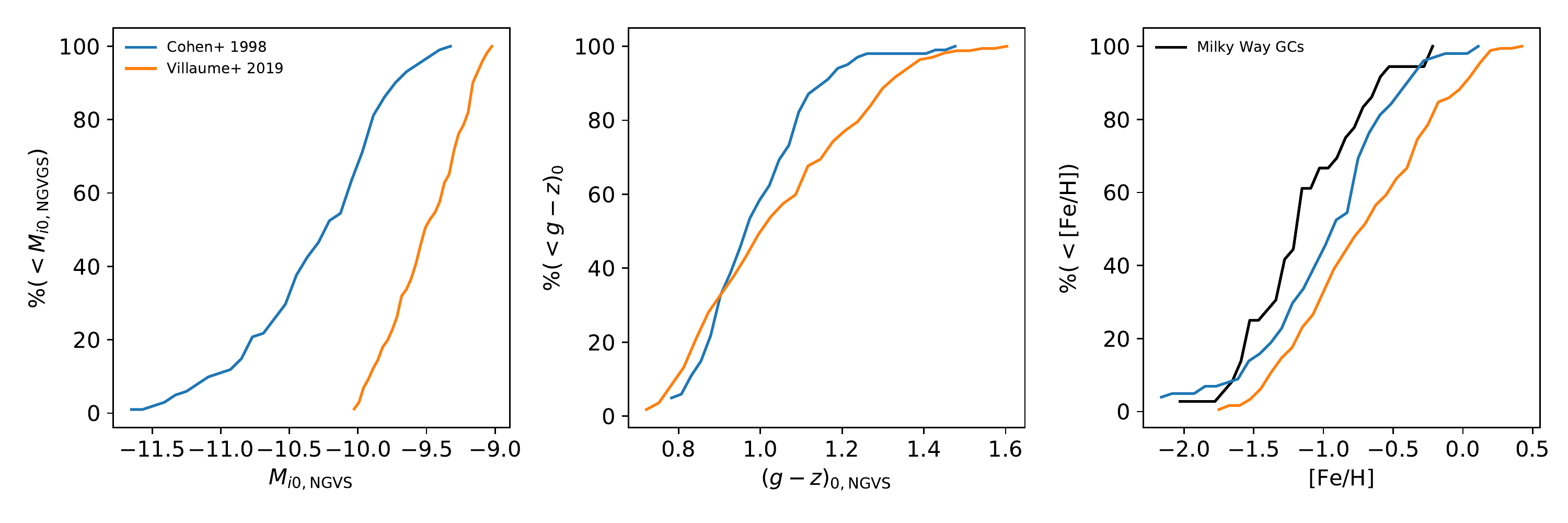}
\caption{Left: Comparing the cumulative magnitude functions for the matched \citet{cohen1998} sample and the sample from this work. Middle: Same as left but for ($g-z$). Right: Same as left and middle but for [Fe/H]. Also in the right panel we show the distribution of [Fe/H] values for Milky Way GCs from our full-spectrum fits to the \citet{schiavon2005} data, which shows that the Milky Way GCs are typically more metal-poor than the M87 GCs.}
\label{fig:cohen_comparison}
\end{figure*}

In Figure~\ref{fig:cohen_comparison} we compare our final sample of 177 GCs to the photometry-matched \citet{cohen1998} sample. In the left panel we compare the cumulative brightness distributions of each sample. In the middle panel we compare the NGVS ($g-z$) colors of the two sample. In the right panel we compare the cumulative metallicity distributions of both samples. We see that $\sim  40\%$ of the objects in our sample are fainter than the faintest GC included in the \citet{cohen1998} sample. The range of colors spanned by each sample are similar but the \citet{cohen1998} sample has a different overall distribution than our sample. More importantly, we see that from the way the curves change from color to metallicity that the \citet{cohen1998} color--metallicity relation will be different than ours. Furthermore, the \citet{cohen1998} metallicities are, on the whole, lower than our metallicities. We discuss the nature of this last difference in more detail in Section 4.1.

\subsection{Updated color--metallicity Relationships}
\label{sec:color--metallicity relation}

\begin{figure}
\includegraphics[width=0.5\textwidth]{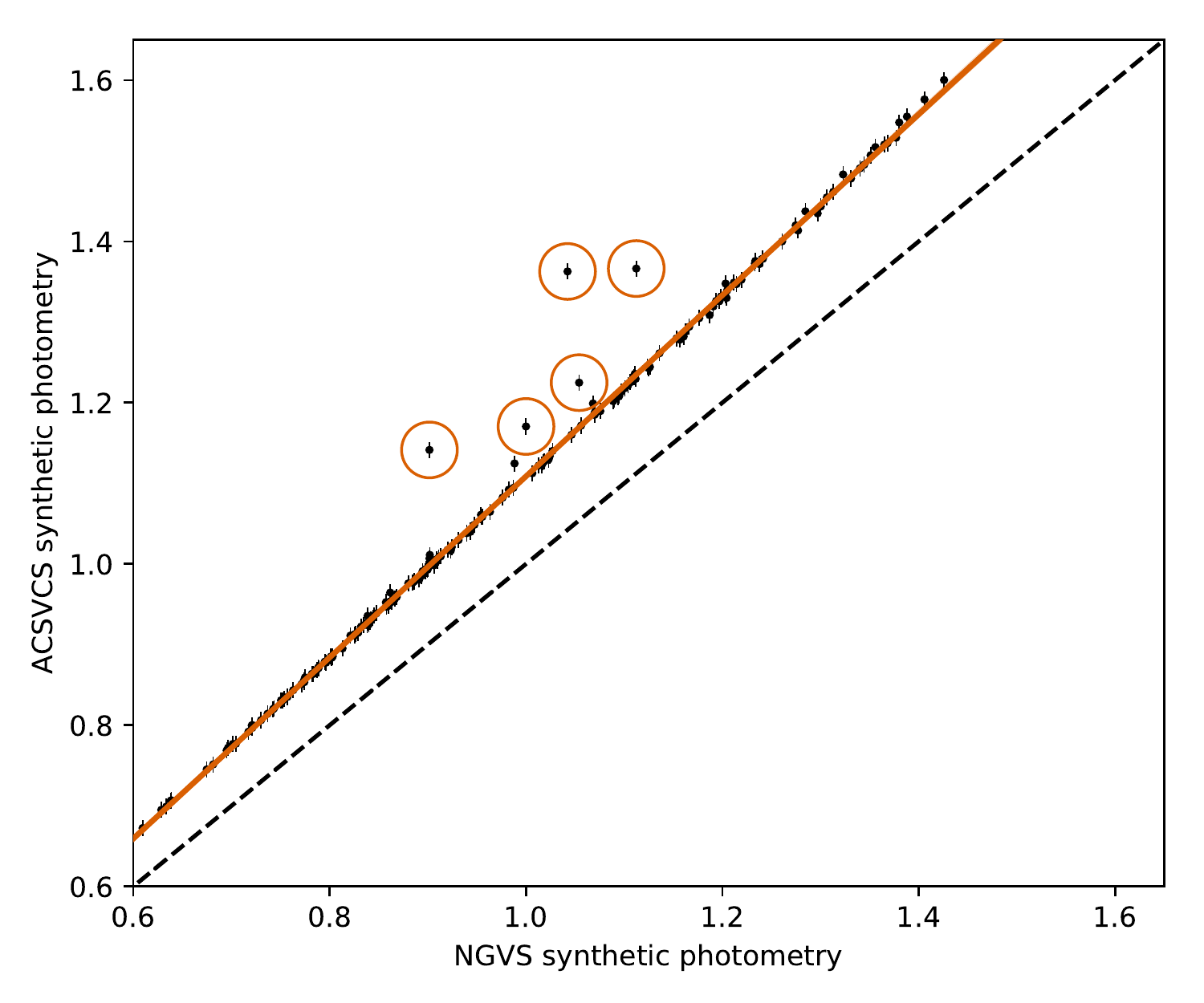}
\caption{Relation between synthetic NGVS and ACSVCS photometry for the spectroscopic sample. Since the two surveys are on slightly different filter systems we present a way to transform colors between each: $(g-z)_{\rm ACSVCS} = 1.123(g-z)_{\rm NGVS} - 0.015$.}
\label{fig:filter_comparison}
\end{figure}

We use two photometric datasets of the M87 GC system: the \citet{oldham2016} catalog of ground based photometry using the NGVS survey data \citep[][]{ferrarese2012} and photometry from the ACS Virgo Cluster Survey (ACSVCS) from \citet{jordan2009}. We use the $g$- and $z$-band filters from each survey but it is important to note that the filters are not identical between the two instruments (see Figure~\ref{fig:filter_comparison}) and so the color--metallicity relationships for the two instruments will be slightly different. 

\begin{figure*}
\includegraphics[width=1\textwidth]{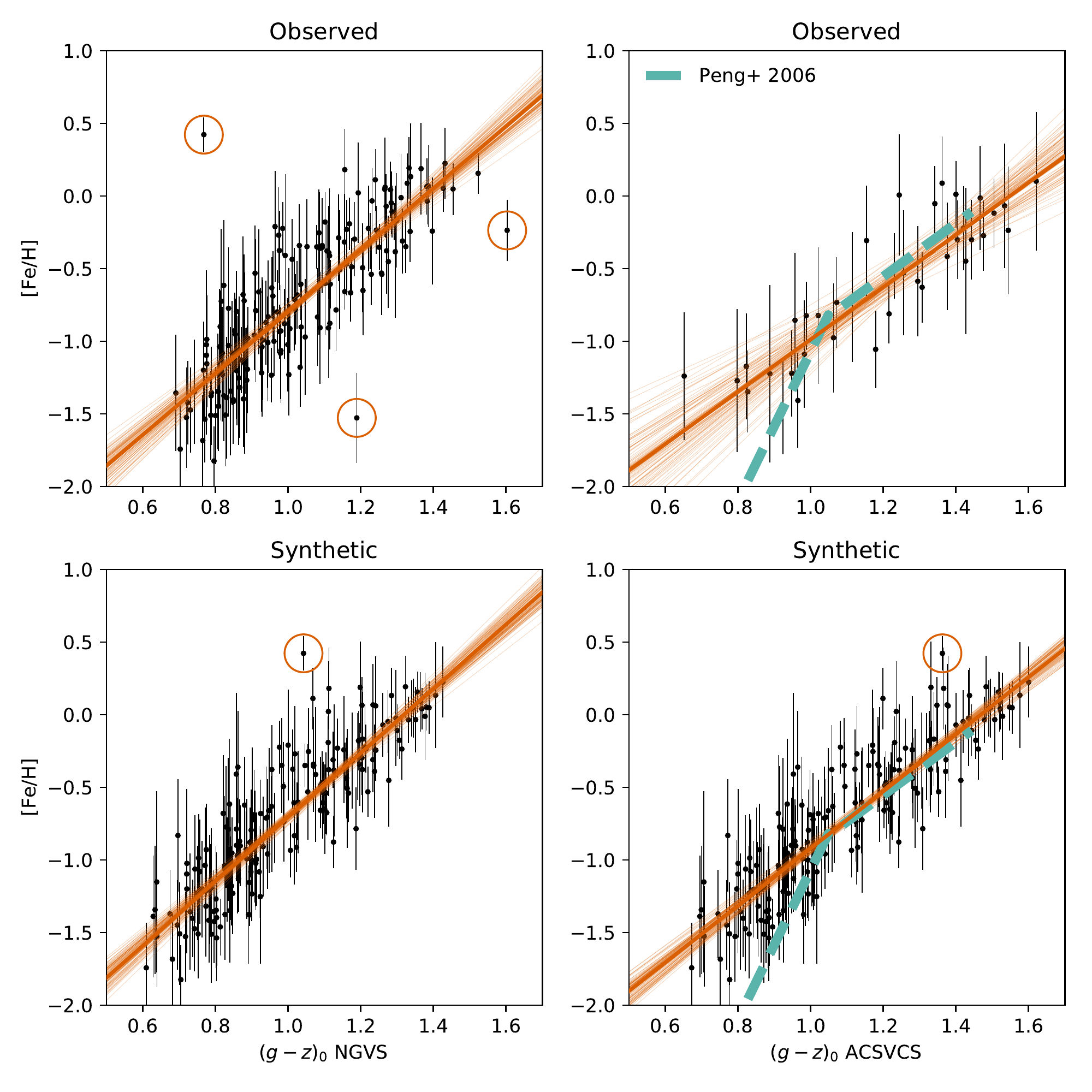}
\caption{(Top-left): Color--metallicity relation using observed NGVS $g-z$ colorsfor the 172 GCs that are in both the spectroscopic sample and NGVS. (Bottom-left): Same as top-left with synthetic colors for all 177 GCs in the spectroscopic sample. (Top-right): Color--metallicity relation using observed ACSVCS colors for the 37 GCs that are in both that and the spectroscopic sample. (Bottom-right): Synthetic color--metallicity relation in the ACSVCS bands for all 177 GCs in the spectroscopic sample. In each panel we show the best-fit line and 100 samples drawn from the posterior distribution by fitting the corresponding data points with a linear model (see text for details). In the right panels we show the \citet{peng2006} relation (dashed green). The regression algorithm detects outliers in the data which are shown in each plot by the red circles.}
\label{fig:color_met}
\end{figure*}

Our sample of 177 spectroscopically-derived [Fe/H] values overlaps with 172 objects from the NGVS catalog but only 37 of the GCs with spectroscopically-derived metallicities overlap with the ACSVCS catalog. To mitigate any problems that might arise from such a sparse sample we leverage the fact that the underlying \texttt{alf} models extend over a wider wavelength range than the LRIS data and are flux calibrated  \citep[see][for discussion]{villaume2017a, conroy2018a}.

We used the flux-calibrated models that correspond to the inferred stellar parameters for each individual GC to compute synthetic photometry for both the ACSVCS and NGVS bandpasses. In Figure~\ref{fig:filter_comparison} we show the relation between the synthetic photometry using the different filter systems. We also show our best-fit line to the data (excluding the outliers marked with the open circles) so that the colors of GCs can be transformed from one system to the other. GCs identified as outliers by the regression model are marked with open circles. The outliers from this relation are just the result of numerical problems for these particular clusters in generating models over the available wavelength range. As can be seen in Figure~\ref{fig:filter_comparison}, the overwhelming majority of the GCs follow a tight relation between the ACSVCS filter system and the NGVS system.

\begin{figure*}
\includegraphics[width=1\textwidth]{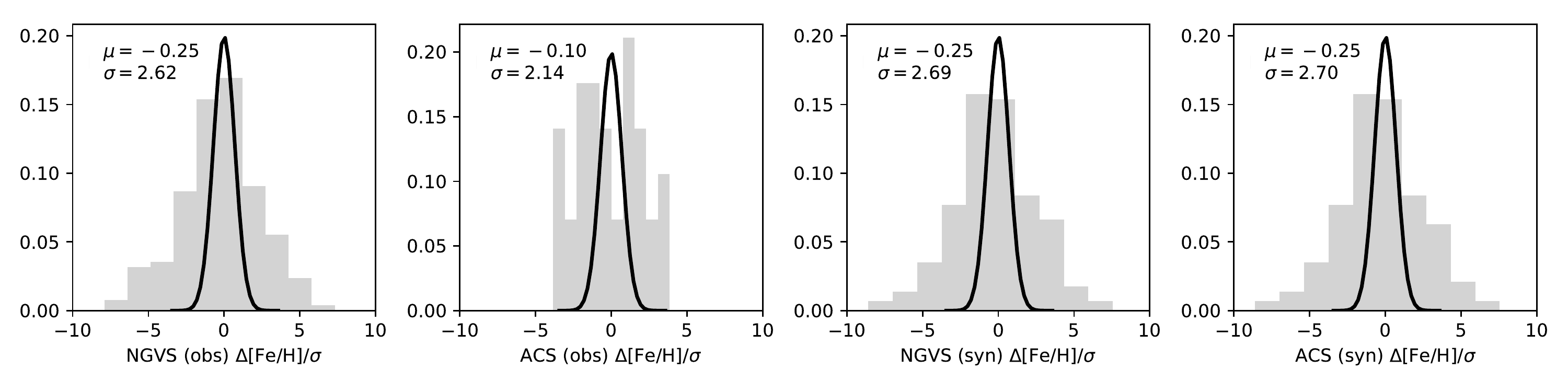}
\caption{Normalized histograms of the residuals between the observed [Fe/H] values and the values predicted by the best-fit color--metallicity relations divided by the observed [Fe/H] uncertainties. We have indicated the mean offset, $\mu$, and standard deviation, $\sigma$ for the distribution of residuals. A Gaussian distribution with $\sigma = 1$ is  also shown.}
\label{fig:residuals}
\end{figure*}

\begin{table}[]
\begin{tabular}{llllll}
           & Slope & $\sigma_{\rm slope}$  & Intercept & $\sigma_{\rm intercept}$  & $\sigma_{\rm residuals}$\\
        \hline
ACSVCS (obs)  &   $1.79$     &  $ 0.25$     &    $-2.77$     &   $0.31$ &  $2.14$               \\
ACSVCS (syn)  &   $1.96$     &  $0.08$      &    $-2.88$     &   $0.10$  &  $2.70$             \\
NGVS (obs)       &   $2.12$     &  $0.12$      &    $-2.92$     &  $0.13$  &  $2.62$            \\
NGVS (syn)       &   $2.20$     &  $0.10$      &    $-2.90 $      &   $0.11$  & $2.69$                
\end{tabular}
\caption{Median values of posterior distributions of best-fit line parameters with standard deviations for each fit. We also show the standard deviation of the residual [Fe/H] distributions,  $\sigma_{\rm residuals}$.}
\label{table:cmr_values}
\end{table}

In Figure~\ref{fig:color_met} we show the color--metallicity relations using the NGVS (left) and ACSVCS (right) photometry for both the observed (top) and  the synthetic (bottom) $g-z$ colors. We fit all four color--metallicity relations using linear regression in a Bayesian framework with outlier pruning and uncertainty weighting \citep[see][for details]{hogg2010} and show the best-fit lines for each relation and 100 samples drawn from the posteriors in each panel (orange lines).

We demonstrate that there is good agreement between the relations using observed and synthetic NGVS photometry. This is important because this assures us of the quality of the synthetic color--metallicity relation for the ACS photometry. The relation using the observed ACSVCS photometry has large uncertainties because of the sparsity of the sample.

Any outliers detected by the fitting algorithm are highlighted by red open circles in each panel. The regression fits do not include those points. Linearity is a good representation of the data in all four cases. We fit the data with a quadratic relation which was not statistically preferred over the linear relation in any case.  In Table~\ref{table:cmr_values} we list the median and standard deviation of slope and intercept values of each relation.

In Figure~\ref{fig:residuals} we show the normalized histograms of the residuals between the observed [Fe/H] values and the values predicted by the best-fit color--metallicity relations divided by the observed [Fe/H] uncertainties. In each panel we show a standard normal distribution and indicate in the legend the measured mean and variance of the residual distribution. The residuals have a larger variance than what is expected from a standard normal distribution. This is likely because the color--metallicity relations have genuine spread since GC systems are an amalgamation of different stellar populations.

In the right panels of Figure~\ref{fig:color_met} we also show the \citet{peng2006} relation. Our relation is consistent with \citet{peng2006} for the red ($g-z > 1.0$) GCs but differs significantly for the blue GCs. We already noted in the previous section that the \citet{cohen1998} metallicities used by \citet{peng2006} are more metal-poor as a whole than the metallicities that we have derived for the M87 GCs. \citet{peng2006} also supplemented their sample with Milky Way GCs. 

\begin{figure*}
\includegraphics[width=1\textwidth]{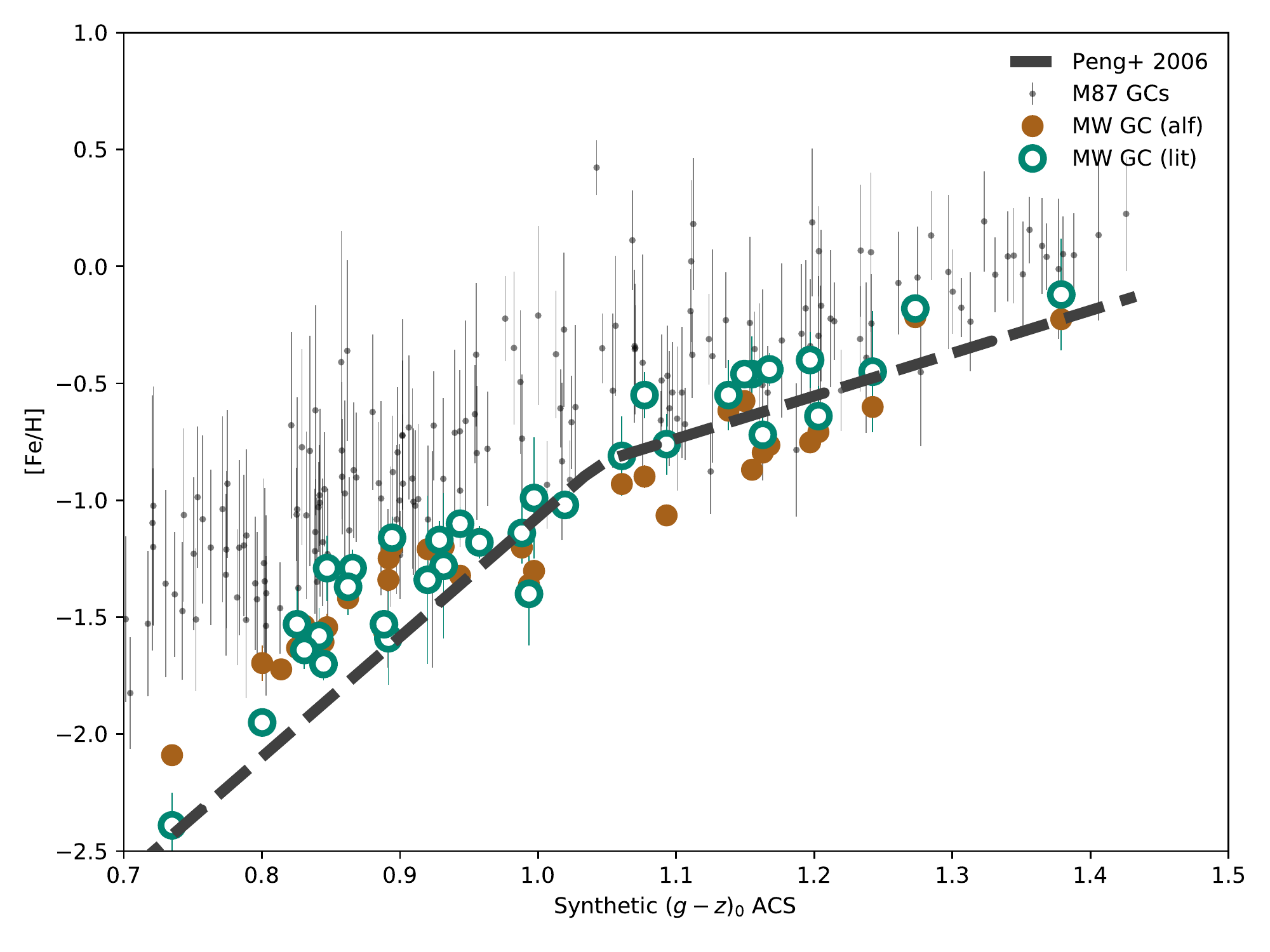}
\caption{We show synthetic ACS $g-z$ color versus metallicity for the M87 clusters (black) and the Milky Way GCs. The inclusion of the MW GCs in the \citet{peng2006} analysis explains much of the discrepancy between our color--metallicity relations. }
\label{fig:mwgc_cmr}
\end{figure*}

To understand how the presence of Milky Way GCs might have affected the color--metallicity relation we look at how the Milky Way GCs compare to the M87 GCs in Figure~\ref{fig:mwgc_cmr}. We generated synthetic photometry for the Milky Way GCs to obtain ACS $g-z$ colors for the clusters. We show the color--metallicity relation using both the [Fe/H] values we derived from our fits to the \citet{schiavon2005} spectroscopy (brown circles) and [Fe/H] values compiled from various literature sources \citep[open green circles]{roediger2014}. We also show the M87 GCs (black points). We show the best-fit lines for the Milky Way GC color--metallicity relation (colored lines) and the \citet{peng2006} relation (dashed black line). In the left panel we show the blue GCs and in the right panel we show the red GCs. 

We see in Figure~\ref{fig:mwgc_cmr} that the blue Milky Way GCs have a different color--metallicity relation than the M87 GCs. The color--metallicity relations for the Milky Way GCs are closer to the \citet{peng2006} relation, which makes sense because it is the Milky Way GCs that drive the blue end of \citet{peng2006} relation. Moreover, \citet{peng2006} used the \citet{harris1996} compilation of Milky Way GC [Fe/H] values and we show that the color--metallicity relation using [Fe/H] values from literature is even closer to the \citet{peng2006} relation than the relation using the spectroscopically derived [Fe/H] values.

\subsection{Metallicity Distributions}

\begin{figure*}
\includegraphics[width=1\textwidth]{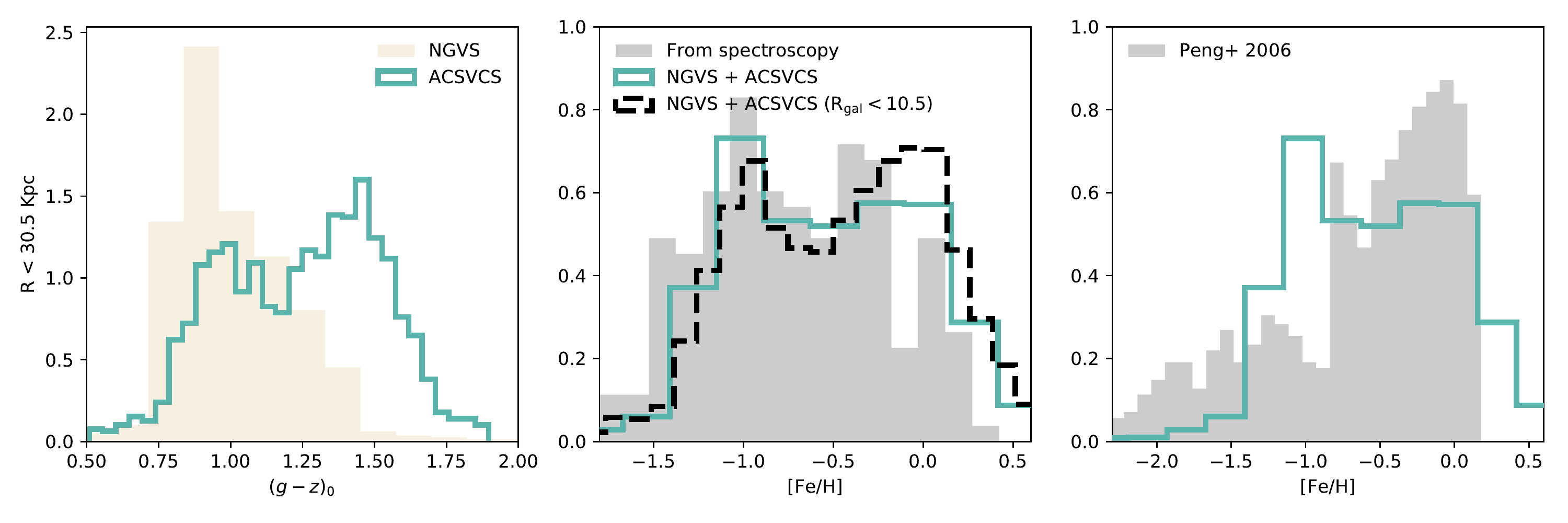}
\caption{(Left): Distributions of the $(g-z)_0$ colors from NGVS (yellow) and ACSVCS (green). The ACSVCS sample is redder and more metal-rich than the NGVS sample on average because it is drawn from a more central region of the galaxy. We limited the NGVS sample to objects within $R_{\rm gal} < 30.5$ kpc to match the footprint of the spectroscopic sample. (Middle): Comparing distributions of metallicity measured from spectroscopy (grey) and from our color--metallicity relationships including both NGVS and ACSVCS photometry where we truncate the sample to $R_{\rm gal} < 30.5$ kpc (green) and  $R_{\rm gal} < 10.5$ kpc (black--dashed). Objects in both samples were removed from the NGVS sample. (Right): Comparing the derived metallicity distributions from NGVS and ACSVCS with the metallicity distribution derived from the \citet{peng2006} relation applied to the ACSVCS colors. The peak of the metal-poor subpopulations are dramatically different, which will affect comparisons to models.}
\label{fig:distributions}
\end{figure*}

In Figure~\ref{fig:distributions} we demonstrate the effect of our new color--metallicity relations on the derived metallicity distributions. In the left panel we compare the NGVS (yellow) and ACSVCS (green) color distributions. For NGVS we only show clusters within $R_{\rm gal} < 30.5$ kpc to match the spatial extent of the spectroscopic dataset. This comparison emphasizes the effect that the spatial extent of the data has on the analysis. In GC systems around massive galaxies, it has been established that the blue GCs begin to dominate further away from the center \citep[e.g.,][]{harris2017a}. The ACSVCS sample only extends to $R_{\rm gal} \sim 13 $ kpc  and we see bimodality clearly in the color distribution for that sample. Meanwhile, the NGVS sample extends more than twice as far out and bimodality gets completely washed out in its color distribution.

In the middle panel we compare the spectroscopically-derived metallicity distribution (grey) with the metallicity distributions derived from the  ACSVCS and NGVS samples using their respective color--metallicity relations for two galactocentric radius cut-offs: $R_{\rm gal} < 30.5$ kpc (green) and  $R_{\rm gal} < 10.5$ kpc (black--dashed). We removed those GCs that are in both samples from the NGVS sample. The photometrically-derived MDF appears to be consistent to the spectroscopically-derived MDF but gives less noisy view of the MDF. The MDF where we truncate at  $R_{\rm gal} = 10.5$ kpc more obviously displays bimodality than the MDF where the sample extends further out.

In the right panel we compare the metallicity distributions derived from the ACSVCS and NGVS colors to the metallicity distribution derived from applying the \citet{peng2006} color--metallicity relation to the ACSVCS data only (grey). We see that the different color--metallicity relations lead to drastically different MDFs. The peak of the metal-poor subpopulation is more metal-rich in MDF established in this work and the dispersions of both subpopulations are very different between the different MDFs.

\begin{figure*}
\includegraphics[width=1\textwidth]{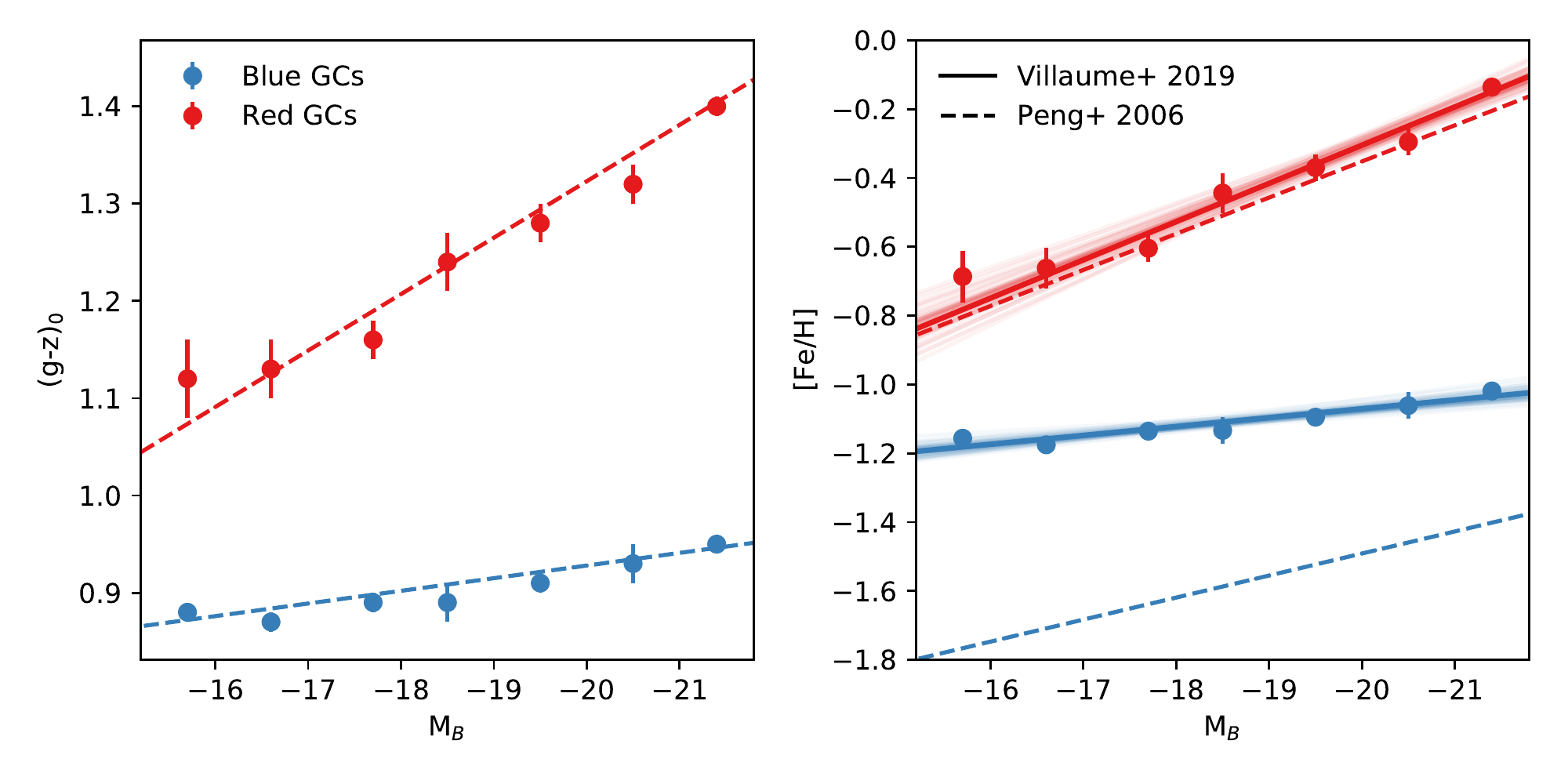}
\caption{(Left) Mean values of the blue and red GC colors as a function of host galaxy luminosity in seven bins of host galaxy magnitude \citep[see][for details]{peng2006}. (Right) Mean metallicities of the blue and red GC populations using the color--metallicity relation determined in this work (solid lines) and the best-fit lines from \citet{peng2006} (dashed lines). The different color--color metallicity established in this work propagates to a dramatically different metal-poor relation.}
\label{fig:p2006_comparison}
\end{figure*}

In Figure~\ref{fig:p2006_comparison} we see the importance of the new color--metallicity relations derived in this work. In the left panel we show the mean values of the blue and red GC populations as a function of host galaxy luminosity in seven bins of host galaxy magnitude for the Virgo Cluster galaxies included in the \citet{peng2006} analysis. In the right panel we have used the color--metallicity relation determined in this paper to transform the mean colors established in \citet{peng2006} into mean metallicities. We derived uncertainties for the metallicity values by doing Monte Carlo sampling of the color--metallicity relation using the color uncertainties. 

We show the linear fit to the new relations in the solid lines. We show the relations \citet{peng2006} determined as dashed lines. As we would expect from the previous results, the new relation between host galaxy luminosity and mean metallicity for the metal-rich GCs is similar to the \citet{peng2006} result but the relation for the metal-poor GCs is shallower and more metal rich than the \citet{peng2006} result.

\section{Discussion}
\label{sec:discussion}

\subsection{Which Metallicity is it Anyway?} 

The difference between our and the \citet{peng2006} color--metallicity relationship is substantial for the blue GCs. We can understand this difference by examining the origin of the [Fe/H] values \citet{peng2006} used in their analysis. First, the Milky Way GCs make up the majority of the blue GCs used in the \citet{peng2006} sample. We demonstrated in Figure~\ref{fig:cohen_comparison} that the Milky Way GCs are more metal-poor than the M87 GCs. In Figure~\ref{fig:mwgc_cmr} we show that, using both literature [Fe/H] values and [Fe/H] values derived from full-spectrum fitting, the Milky Way GCs have a different color--metallicity relation than the M87 GCs. The closeness of the \citet{peng2006} relation in the blue to the Milky Way GC relation is highly suggestive that the presence of the Milky Way GCs is driving and biasing the relation in the blue for \citet{peng2006}.

Second, we show in Figure~\ref{fig:cohen_comparison} that even though the GCs in our sample and the \citet{cohen1998} sample span a similar color range, the \citet{cohen1998} metallicity values are systemically lower than the metallicities we derive. There are no GCs that are shared between the \citet{cohen1998} sample and our sample but we can understand the differences between the two by bearing in mind two related facts: the fitting functions that underlie the \citet{worthey1994} models are not well-calibrated at high metallicities and the \citet{cohen1998} metallicities are placed on the \citet{zinn1985} metallicity scale which is set by Milky Way GCs.

The former was discussed in \citet{cohen2003} as a serious concern. \citet{cohen2003} redid the [Fe/H] determinations of the M87 GCs from \citet{cohen1998} by extrapolating the models to higher metallicity by assuming that the indices are on the damping part of the curve of growth. This affected five M87 GCs in their sample. We are, to be clear, using the \citet{cohen1998} metallicities in this work as \citet{peng2006} did.

For the latter, \citet{cohen1998} noted that from their qualitative analysis of the line indices of both the Milky Way and M87 GCs, the M87 GCs have a metal-rich tail that extends to significantly higher metallicities than the Milky Way GCs, which we confirm. The relation they use to scale the \citet{worthey1994} models to the Milky Way GCs is ${\rm [Fe/H]_Z} = 0.760 \times {\rm [Fe/H]_W} - 0.265$ which would lower the overall metallicity  of their sample. Overall, we see that the \citet{peng2006} relation is yoked to the Milky Way GCs in both explicit and implicit ways. The [Fe/H] values that we present in this work come from the underlying isochrones \citep{choi2016} and the underlying stellar library \citep{villaume2017a}. While the stellar library consists of stars from the Milky Way, there is not a Milky Way specific trend in [Fe/H] that we need to correct like we would for $\alpha$ elements \citep[][]{tripicco1995}.

We also find that the color--metallicity relation differs between the Milky Way and M87, especially near the blue end ($g-z \lesssim 1.0$). We defer an in-depth examination of the physical cause of this difference to later work but we speculate that it might be age differences between the two GC populations. If the M87 GCs were younger than the Milky Way GCs, they would appear bluer at the same metallicities. We used simple stellar population (SSP) synthesis models to examine how age affects color at fixed metallicity (in this case, [Fe/H]$ = -1.5$) and found that the metal-poor M87 GCs would have to be about 4 Gyr younger than the Milky Way GCs to explain the color difference. We also cannot rule out the possible effects that $\alpha$ elements or the morphology of the blue horizontal branch have on the color.

\subsection{Bimodality}

Bimodality of GC systems has been the dominant paradigm in which extragalactic GC studies have been conducted over the past 30 years. In this paper we defer quantitative analysis of the subpopulations of the GC system around M87 to a forthcoming paper on the subject. This is to more appropriately address the complexities around the topic that have been raised recently. Even with just the M87 system, consensus has yet to be reached on the number of subpopulations that make up the system \citep[e.g.,][]{strader2011, agnello2014b, oldham2016}. With that being said, there are still some things worth pointing out. 

First, \citet{cohen1998} detected bimodality in M87 only after excluding the metal-rich tail from their analysis. \citet{usher2012} speculated that the lack of convincing proof for bimodality from \citet{cohen1998} was a result of their typically bright targets. Since \citet{cohen1998} we have become aware of the blue-tilt phenomenon as well as liminal objects like ultra-compact dwarfs that could contaminate populations of bright canonical GCs \citep[e.g.,][]{usher2012, villaume2017b}.

Second, we take advantage of obtaining color--metallicity relations using both the NGVS and ACSVCS datasets by converting both into metallicity and combining the data sets. The ACSVCS data probe the very inner region of the M87 GC system while the NGVS data extends further out. We see that the color-converted MDF is consistent with the spectroscopically determined MDF. Furthermore, bimodality can be seen visually from the MDFs, especially when only GCs within R$_{\rm gal} < 10.5$ kpc are included.
 
The M87 system consists of a huge number of GCs that represent the culmination of a complex history. Previous analyses \citep[e.g.,][]{strader2011, romanowsky2012} indicated that the GCs in the outer halo behave differently and are dominated by blue/metal-poor GCs. As mentioned previously, a paper specifically addressing the subpopulations and their characteristics will follow this paper.

Third, the Yonsei Evolutionary Population Synthesis (YEPS) models have been used to argue that most bimodal color distributions reflect a truly unimodal underlying metallicity distribution because of the inclusion of hot horizontal branch stars \citep[][]{yoon2006}. The approach of this group is different from the one typically taken, where spectroscopic observations of individual GCs are modeled with SPS models. Instead, the YEPS group transforms the color distributions of GC systems to metallicity distributions using synthetic color--metallicity relations generated from the YEPS models.

The results from the synthetic color--metallicity relation method \citep[][]{yoon2006, lee2019} and the direct spectroscopic modeling \citep[e.g.,][]{alves-brito2011, usher2012, brodie2012} method continue to be at odds. The results we find in this work are consistent with other studies that have directly modeled spectroscopy of individual GCs. Beyond the final results there are few points of comparison between the two methods. However, we note that from our work with Milky Way GCs we know that the presence of hot horizontal branch stars affects our ability to measure accurate ages from integrated light but not metallicity \citep[see Figure 15 in][for reference]{conroy2018a}. We therefore do not have a reason to doubt our metallicity measurements for the M87 GCs, even with the possible presence of GCs with prominent hot horizontal branches.

Fourth, it is important to note that our MDFs, both from the purely spectroscopic sample and the sample converting NGVS and ACSVCS photometry, differ significantly from the MDF computed from the \citet{peng2006} relation. The peaks and widths of the distributions are different. These quantities are crucial for making quantitative comparisons to theoretical models of GC system formation, and thus, of galaxy formation. In recent years modern theoretical galaxy formation models have emerged with the E-MOSAICS simulations \citep[][]{pfeffer2018} for Milky Way-type galaxies, and alternatively with \citet{choksi2018} specifically trying to recreate the observed properties of the Virgo Cluster galaxies. These models take divergent approaches: E-MOSAICS adds models describing the formation and evolution of star clusters into the EAGLE galaxy formation simulations, while \citet{choksi2018} uses semi-analytic models of merger histories. They also take different approaches to the role GC destruction plays in our understanding of $z\sim 0$ GC systems. Accuracy and credible uncertainties in  the physical characteristics derived from observables are crucial for moving forward with constraining galaxy formation theories based on GCs.

\subsection{Implications for GC and Galaxy Formation}

We have derived a new galaxy luminosity--GC metallicity relation separately for the blue and red GCs in the Virgo galaxies included in \citet{peng2006} (Figure~\ref{fig:p2006_comparison}). The difference in our new color--metallicity relation is two-fold: the metal-poor GCs now correlate with galaxy luminosity {\it less strongly} than previously measured, and the metal-poor GCs are more metal-rich than what \citet{peng2006} determined.

\citet{larsen2001} were the first to assess the relationship between GC {\it subpopulation} metallicity and galaxy luminosity with a homogeneously acquired sample. Then \citet{strader2004} combined elliptical galaxy data from a variety of sources \citep[][]{larsen2001, kundu2001a, kundu2001b} with data from spiral galaxies \citep[][]{harris1996, barmby2000} to look at just the metal-poor GCs. Most recently \citet{peng2006} determined this relationship for the Virgo Cluster galaxies. Like \citet{larsen2001}, \citet{peng2006} found shallower slopes for the metal-poor GCs relative to the metal-rich GCs. There is remarkable similarity between the slopes that \citet{larsen2001}, \citet{strader2004}, and \citet{peng2006} found for the metal-poor GCs. 

We already know that the difference between our relation and the relation from \citet{peng2006} is due to the color--metallicity relation. What about the difference with \citet{strader2004}? \citet{strader2004} used the \citet{barmby2000} color--metallicity relation based on a sample of M31 GCs. \citet{barmby2000} noted that the M31 color--metallicity relation is similar to the Milky Way relation. This raises the likelihood that it is not an appropriate way to convert colors to metallicities for the early-type galaxies included in the \citet{strader2004} sample. The similarity in slopes between \citet{strader2004} and \citet{peng2006} might be an artifact of the similar source of their respective color--metallicity relations.

To explain the correlation between galaxy luminosity and blue GC metallicity \citet{strader2005} and \citet{brodie2006} invoked the concept of ``biasing'', also introduced in the context of Milky Way stellar halo assembly by \citet{robertson2005}. In short, the progenitor satellites that now constitute the stellar halos of massive galaxies were more metal-rich, at fixed mass, than present day satellites. In the light of the new, much weaker correlation, this needs to be reassessed. The new correlation could indicate that biasing is not as strong as an effect as once thought. Put another way, the new correlation suggests that metal-poor GCs formed irrespective of their host galaxies.
 
The change in metallicity intercept for the metal-poor GCs on this relation has implications for their formation epoch. \citet{forbes2015} evolved the galaxy mass--GC metallicity relation through redshift to determine bulk ages of the GCs belonging to the galaxies in the SLUGGS survey \citep[e.g.,][]{usher2012}. In their model, higher metallicities indicate younger ages and/or more massive hosts. The slopes of their metal-poor and metal-rich relations are not totally consistent with what we present in this work, but the intercepts are roughly similar. Following the logic of \citet{forbes2015}, the nearly constant slope we find for the metal-poor GCs as a function of galaxy luminosity indicates that the metal-poor GCs in the Virgo Cluster formed at nearly the same time. The correlation between the metal-rich GC [Fe/H] values and host galaxy luminosity indicates that the metal-rich GCs around the giant galaxies formed more recently than the metal-rich GCs around the dwarf galaxies. The increase in metallicity for the metal-poor GCs may also help ease the tension between simulated and observational results as discussed in the Introduction, if it indicates that the GCs formed in more massive satellites.

It is important to note the crucial underlying caveat of Figure~\ref{fig:p2006_comparison} -- that the color--metallicity relation we developed for M87 is applicable to the other Virgo Cluster galaxies in the \citet{peng2006} analysis. This is probably not a good assumption, particularly in light of the \citet{powalka2016a} results which showed that color--color relations in the NGVS sample depend on environment, with colors on the whole becoming bluer with increased radial distance from M87 and that GCs  $>200$ kpc from M87 have color--color distributions similar to those of the Milky Way. Unfortunately, \citet{powalka2016a} also showed that mass is not the driving factor in these differences so we cannot make a simple correction to Figure~\ref{fig:p2006_comparison}. More detailed spectroscopy of lower-mass systems in the Virgo Cluster is ultimately needed.

\section{Summary}

\begin{itemize}

\item We have fitted a spectroscopic sample of GCs around M87 with full-spectrum SPS models and obtained [Fe/H] for 177 GCs. We demonstrate that the metallicity values we derive are systematically higher-metallicity than previous spectroscopic studies. We attribute this difference to the limitations of the previously-used \citet{worthey1994} SPS models and because the previously determined metallicity values were scaled to match the Milky Way GCs, which are, as a whole, lower in metallicity than the M87 GCs.
\item Using synthetic photometry from flux-calibrated stellar population models we determine a transformation between the NGVS and ACSVCS photometric systems: $(g-z)_{\rm ACSVCS} = 1.123(g-z)_{\rm NGVS} - 0.015$.
\item We derived new color--metallicity relations using both NGVS and ACSVCS $g-z$ colors. Our ACSVCS color--metallicity relation differs significantly for the blue GCs from the previously published color--metallicity relation using the ACS filters. This is because we find the relation for the Milky Way GCs to be significantly different than the relation for the M87 GCs. We discuss the necessary age difference needed to explain this result, but previous work in colors of Virgo Cluster GCs suggested that there is some environmental effect on chemical abundance patterns.%We do not comment on the nature of the difference in color--metallicity relations among galaxies, but previous work in colors of Virgo Cluster GCs suggests that there is some environmental effect on chemical abundance patterns.
\item While we advocate that color--metallicity relations be confirmed with spectroscopic follow-up for individual galaxies, we assume that in this respect the Virgo cluster galaxies are similar to one another and as a result we find a shallower galaxy luminosity-GC metallicity relation for the blue GCs than previous studies. This could either indicate that progenitor satellites were less massive than previously thought, or the properties of metal-poor GCs are not as dependent on their present-day host galaxy as metal-rich GCs. 

\end{itemize}

\bibliography{/Users/alexa/gradients/Paper/references}{}

\begin{thebibliography}{65}
\expandafter\ifx\csname natexlab\endcsname\relax\def\natexlab#1{#1}\fi
\expandafter\ifx\csname bibnamefont\endcsname\relax
  \def\bibnamefont#1{#1}\fi
\expandafter\ifx\csname bibfnamefont\endcsname\relax
  \def\bibfnamefont#1{#1}\fi
\expandafter\ifx\csname citenamefont\endcsname\relax
  \def\citenamefont#1{#1}\fi
\expandafter\ifx\csname url\endcsname\relax
  \def\url#1{\texttt{#1}}\fi
\expandafter\ifx\csname urlprefix\endcsname\relax\def\urlprefix{URL }\fi
\providecommand{\bibinfo}[2]{#2}
\providecommand{\eprint}[2][]{\url{#2}}

\bibitem[{\citenamefont{{Pillepich} et~al.}(2018)\citenamefont{{Pillepich},
  {Nelson}, {Hernquist}, {Springel}, {Pakmor}, {Torrey}, {Weinberger}, {Genel},
  {Naiman}, {Marinacci} et~al.}}]{pillepich2018}
\bibinfo{author}{\bibfnamefont{A.}~\bibnamefont{{Pillepich}}},
  \bibinfo{author}{\bibfnamefont{D.}~\bibnamefont{{Nelson}}},
  \bibinfo{author}{\bibfnamefont{L.}~\bibnamefont{{Hernquist}}},
  \bibinfo{author}{\bibfnamefont{V.}~\bibnamefont{{Springel}}},
  \bibinfo{author}{\bibfnamefont{R.}~\bibnamefont{{Pakmor}}},
  \bibinfo{author}{\bibfnamefont{P.}~\bibnamefont{{Torrey}}},
  \bibinfo{author}{\bibfnamefont{R.}~\bibnamefont{{Weinberger}}},
  \bibinfo{author}{\bibfnamefont{S.}~\bibnamefont{{Genel}}},
  \bibinfo{author}{\bibfnamefont{J.~P.} \bibnamefont{{Naiman}}},
  \bibinfo{author}{\bibfnamefont{F.}~\bibnamefont{{Marinacci}}},
  \bibnamefont{et~al.}, \bibinfo{journal}{\mnras}
  \textbf{\bibinfo{volume}{475}}, \bibinfo{pages}{648} (\bibinfo{year}{2018}),
  \eprint{1707.03406}.

\bibitem[{\citenamefont{{Forbes} et~al.}(2015)\citenamefont{{Forbes},
  {Pastorello}, {Romanowsky}, {Usher}, {Brodie}, and {Strader}}}]{forbes2015}
\bibinfo{author}{\bibfnamefont{D.~A.} \bibnamefont{{Forbes}}},
  \bibinfo{author}{\bibfnamefont{N.}~\bibnamefont{{Pastorello}}},
  \bibinfo{author}{\bibfnamefont{A.~J.} \bibnamefont{{Romanowsky}}},
  \bibinfo{author}{\bibfnamefont{C.}~\bibnamefont{{Usher}}},
  \bibinfo{author}{\bibfnamefont{J.~P.} \bibnamefont{{Brodie}}},
  \bibnamefont{and}
  \bibinfo{author}{\bibfnamefont{J.}~\bibnamefont{{Strader}}},
  \bibinfo{journal}{\mnras} \textbf{\bibinfo{volume}{452}},
  \bibinfo{pages}{1045} (\bibinfo{year}{2015}), \eprint{1506.06820}.

\bibitem[{\citenamefont{{Brodie} and {Strader}}(2006)}]{brodie2006}
\bibinfo{author}{\bibfnamefont{J.~P.} \bibnamefont{{Brodie}}} \bibnamefont{and}
  \bibinfo{author}{\bibfnamefont{J.}~\bibnamefont{{Strader}}},
  \bibinfo{journal}{Annual Review of Astronomy and Astrophysics}
  \textbf{\bibinfo{volume}{44}}, \bibinfo{pages}{193} (\bibinfo{year}{2006}),
  \eprint{astro-ph/0602601}.

\bibitem[{\citenamefont{{van den Bergh}}(1975)}]{bergh1975}
\bibinfo{author}{\bibfnamefont{S.}~\bibnamefont{{van den Bergh}}},
  \bibinfo{journal}{Annual Review of Astronomy and Astrophysics}
  \textbf{\bibinfo{volume}{13}}, \bibinfo{pages}{217} (\bibinfo{year}{1975}).

\bibitem[{\citenamefont{{Brodie} and {Huchra}}(1991)}]{brodie1991}
\bibinfo{author}{\bibfnamefont{J.~P.} \bibnamefont{{Brodie}}} \bibnamefont{and}
  \bibinfo{author}{\bibfnamefont{J.~P.} \bibnamefont{{Huchra}}},
  \bibinfo{journal}{\apj} \textbf{\bibinfo{volume}{379}}, \bibinfo{pages}{157}
  (\bibinfo{year}{1991}).

\bibitem[{\citenamefont{{Gebhardt} and {Kissler-Patig}}(1999)}]{gebhardt1999}
\bibinfo{author}{\bibfnamefont{K.}~\bibnamefont{{Gebhardt}}} \bibnamefont{and}
  \bibinfo{author}{\bibfnamefont{M.}~\bibnamefont{{Kissler-Patig}}},
  \bibinfo{journal}{AJ} \textbf{\bibinfo{volume}{118}}, \bibinfo{pages}{1526}
  (\bibinfo{year}{1999}), \eprint{astro-ph/9906499}.

\bibitem[{\citenamefont{{Kundu} and
  {Whitmore}}(2001{\natexlab{a}})}]{kundu2001a}
\bibinfo{author}{\bibfnamefont{A.}~\bibnamefont{{Kundu}}} \bibnamefont{and}
  \bibinfo{author}{\bibfnamefont{B.~C.} \bibnamefont{{Whitmore}}},
  \bibinfo{journal}{\aj} \textbf{\bibinfo{volume}{121}}, \bibinfo{pages}{2950}
  (\bibinfo{year}{2001}{\natexlab{a}}), \eprint{astro-ph/0103021}.

\bibitem[{\citenamefont{{Larsen} et~al.}(2001)\citenamefont{{Larsen}, {Brodie},
  {Huchra}, {Forbes}, and {Grillmair}}}]{larsen2001}
\bibinfo{author}{\bibfnamefont{S.~S.} \bibnamefont{{Larsen}}},
  \bibinfo{author}{\bibfnamefont{J.~P.} \bibnamefont{{Brodie}}},
  \bibinfo{author}{\bibfnamefont{J.~P.} \bibnamefont{{Huchra}}},
  \bibinfo{author}{\bibfnamefont{D.~A.} \bibnamefont{{Forbes}}},
  \bibnamefont{and} \bibinfo{author}{\bibfnamefont{C.~J.}
  \bibnamefont{{Grillmair}}}, \bibinfo{journal}{\aj}
  \textbf{\bibinfo{volume}{121}}, \bibinfo{pages}{2974} (\bibinfo{year}{2001}),
  \eprint{astro-ph/0102374}.

\bibitem[{\citenamefont{{C{\^o}t{\'e}}
  et~al.}(2002)\citenamefont{{C{\^o}t{\'e}}, {West}, and {Marzke}}}]{cote2002}
\bibinfo{author}{\bibfnamefont{P.}~\bibnamefont{{C{\^o}t{\'e}}}},
  \bibinfo{author}{\bibfnamefont{M.~J.} \bibnamefont{{West}}},
  \bibnamefont{and} \bibinfo{author}{\bibfnamefont{R.~O.}
  \bibnamefont{{Marzke}}}, \bibinfo{journal}{\apj}
  \textbf{\bibinfo{volume}{567}}, \bibinfo{pages}{853} (\bibinfo{year}{2002}),
  \eprint{astro-ph/0111388}.

\bibitem[{\citenamefont{{Strader} et~al.}(2005)\citenamefont{{Strader},
  {Brodie}, {Cenarro}, {Beasley}, and {Forbes}}}]{strader2005}
\bibinfo{author}{\bibfnamefont{J.}~\bibnamefont{{Strader}}},
  \bibinfo{author}{\bibfnamefont{J.~P.} \bibnamefont{{Brodie}}},
  \bibinfo{author}{\bibfnamefont{A.~J.} \bibnamefont{{Cenarro}}},
  \bibinfo{author}{\bibfnamefont{M.~A.} \bibnamefont{{Beasley}}},
  \bibnamefont{and} \bibinfo{author}{\bibfnamefont{D.~A.}
  \bibnamefont{{Forbes}}}, \bibinfo{journal}{\aj}
  \textbf{\bibinfo{volume}{130}}, \bibinfo{pages}{1315} (\bibinfo{year}{2005}),
  \eprint{astro-ph/0506289}.

\bibitem[{\citenamefont{{Rhode} et~al.}(2005)\citenamefont{{Rhode}, {Zepf}, and
  {Santos}}}]{rhode2005}
\bibinfo{author}{\bibfnamefont{K.~L.} \bibnamefont{{Rhode}}},
  \bibinfo{author}{\bibfnamefont{S.~E.} \bibnamefont{{Zepf}}},
  \bibnamefont{and} \bibinfo{author}{\bibfnamefont{M.~R.}
  \bibnamefont{{Santos}}}, \bibinfo{journal}{\apj}
  \textbf{\bibinfo{volume}{630}}, \bibinfo{pages}{L21} (\bibinfo{year}{2005}),
  \eprint{astro-ph/0507551}.

\bibitem[{\citenamefont{{Li} and {Gnedin}}(2014)}]{li2014}
\bibinfo{author}{\bibfnamefont{H.}~\bibnamefont{{Li}}} \bibnamefont{and}
  \bibinfo{author}{\bibfnamefont{O.~Y.} \bibnamefont{{Gnedin}}},
  \bibinfo{journal}{\apj} \textbf{\bibinfo{volume}{796}}, \bibinfo{eid}{10}
  (\bibinfo{year}{2014}), \eprint{1405.0763}.

\bibitem[{\citenamefont{{Harris} et~al.}(2017)\citenamefont{{Harris},
  {Ciccone}, {Eadie}, {Gnedin}, {Geisler}, {Rothberg}, and
  {Bailin}}}]{harris2017a}
\bibinfo{author}{\bibfnamefont{W.~E.} \bibnamefont{{Harris}}},
  \bibinfo{author}{\bibfnamefont{S.~M.} \bibnamefont{{Ciccone}}},
  \bibinfo{author}{\bibfnamefont{G.~M.} \bibnamefont{{Eadie}}},
  \bibinfo{author}{\bibfnamefont{O.~Y.} \bibnamefont{{Gnedin}}},
  \bibinfo{author}{\bibfnamefont{D.}~\bibnamefont{{Geisler}}},
  \bibinfo{author}{\bibfnamefont{B.}~\bibnamefont{{Rothberg}}},
  \bibnamefont{and} \bibinfo{author}{\bibfnamefont{J.}~\bibnamefont{{Bailin}}},
  \bibinfo{journal}{\apj} \textbf{\bibinfo{volume}{835}}, \bibinfo{eid}{101}
  (\bibinfo{year}{2017}), \eprint{1612.08089}.

\bibitem[{\citenamefont{{Harris} et~al.}(2006)\citenamefont{{Harris},
  {Whitmore}, {Karakla}, {Oko{\'n}}, {Baum}, {Hanes}, and
  {Kavelaars}}}]{harris2006}
\bibinfo{author}{\bibfnamefont{W.~E.} \bibnamefont{{Harris}}},
  \bibinfo{author}{\bibfnamefont{B.~C.} \bibnamefont{{Whitmore}}},
  \bibinfo{author}{\bibfnamefont{D.}~\bibnamefont{{Karakla}}},
  \bibinfo{author}{\bibfnamefont{W.}~\bibnamefont{{Oko{\'n}}}},
  \bibinfo{author}{\bibfnamefont{W.~A.} \bibnamefont{{Baum}}},
  \bibinfo{author}{\bibfnamefont{D.~A.} \bibnamefont{{Hanes}}},
  \bibnamefont{and} \bibinfo{author}{\bibfnamefont{J.~J.}
  \bibnamefont{{Kavelaars}}}, \bibinfo{journal}{\apj}
  \textbf{\bibinfo{volume}{636}}, \bibinfo{pages}{90} (\bibinfo{year}{2006}),
  \eprint{astro-ph/0508195}.

\bibitem[{\citenamefont{{Peng} et~al.}(2006)\citenamefont{{Peng}, {Jord{\'a}n},
  {C{\^o}t{\'e}}, {Blakeslee}, {Ferrarese}, {Mei}, {West}, {Merritt},
  {Milosavljevi{\'c}}, and {Tonry}}}]{peng2006}
\bibinfo{author}{\bibfnamefont{E.~W.} \bibnamefont{{Peng}}},
  \bibinfo{author}{\bibfnamefont{A.}~\bibnamefont{{Jord{\'a}n}}},
  \bibinfo{author}{\bibfnamefont{P.}~\bibnamefont{{C{\^o}t{\'e}}}},
  \bibinfo{author}{\bibfnamefont{J.~P.} \bibnamefont{{Blakeslee}}},
  \bibinfo{author}{\bibfnamefont{L.}~\bibnamefont{{Ferrarese}}},
  \bibinfo{author}{\bibfnamefont{S.}~\bibnamefont{{Mei}}},
  \bibinfo{author}{\bibfnamefont{M.~J.} \bibnamefont{{West}}},
  \bibinfo{author}{\bibfnamefont{D.}~\bibnamefont{{Merritt}}},
  \bibinfo{author}{\bibfnamefont{M.}~\bibnamefont{{Milosavljevi{\'c}}}},
  \bibnamefont{and} \bibinfo{author}{\bibfnamefont{J.~L.}
  \bibnamefont{{Tonry}}}, \bibinfo{journal}{\apj}
  \textbf{\bibinfo{volume}{639}}, \bibinfo{pages}{95} (\bibinfo{year}{2006}),
  \eprint{astro-ph/0509654}.

\bibitem[{\citenamefont{{Jord{\'a}n} et~al.}(2004)\citenamefont{{Jord{\'a}n},
  {Blakeslee}, {Peng}, {Mei}, {C{\^o}t{\'e}}, {Ferrarese}, {Tonry}, {Merritt},
  {Milosavljevi{\'c}}, and {West}}}]{jordan2004}
\bibinfo{author}{\bibfnamefont{A.}~\bibnamefont{{Jord{\'a}n}}},
  \bibinfo{author}{\bibfnamefont{J.~P.} \bibnamefont{{Blakeslee}}},
  \bibinfo{author}{\bibfnamefont{E.~W.} \bibnamefont{{Peng}}},
  \bibinfo{author}{\bibfnamefont{S.}~\bibnamefont{{Mei}}},
  \bibinfo{author}{\bibfnamefont{P.}~\bibnamefont{{C{\^o}t{\'e}}}},
  \bibinfo{author}{\bibfnamefont{L.}~\bibnamefont{{Ferrarese}}},
  \bibinfo{author}{\bibfnamefont{J.~L.} \bibnamefont{{Tonry}}},
  \bibinfo{author}{\bibfnamefont{D.}~\bibnamefont{{Merritt}}},
  \bibinfo{author}{\bibfnamefont{M.}~\bibnamefont{{Milosavljevi{\'c}}}},
  \bibnamefont{and} \bibinfo{author}{\bibfnamefont{M.~J.}
  \bibnamefont{{West}}}, \bibinfo{journal}{The Astrophysical Journal Supplement
  Series} \textbf{\bibinfo{volume}{154}}, \bibinfo{pages}{509}
  (\bibinfo{year}{2004}), \eprint{astro-ph/0406433}.

\bibitem[{\citenamefont{{Cohen} et~al.}(1998)\citenamefont{{Cohen},
  {Blakeslee}, and {Ryzhov}}}]{cohen1998}
\bibinfo{author}{\bibfnamefont{J.~G.} \bibnamefont{{Cohen}}},
  \bibinfo{author}{\bibfnamefont{J.~P.} \bibnamefont{{Blakeslee}}},
  \bibnamefont{and} \bibinfo{author}{\bibfnamefont{A.}~\bibnamefont{{Ryzhov}}},
  \bibinfo{journal}{\apj} \textbf{\bibinfo{volume}{496}}, \bibinfo{pages}{808}
  (\bibinfo{year}{1998}), \eprint{astro-ph/9709192}.

\bibitem[{\citenamefont{{Cohen} et~al.}(2003)\citenamefont{{Cohen},
  {Blakeslee}, and {C{\^o}t{\'e}}}}]{cohen2003}
\bibinfo{author}{\bibfnamefont{J.~G.} \bibnamefont{{Cohen}}},
  \bibinfo{author}{\bibfnamefont{J.~P.} \bibnamefont{{Blakeslee}}},
  \bibnamefont{and}
  \bibinfo{author}{\bibfnamefont{P.}~\bibnamefont{{C{\^o}t{\'e}}}},
  \bibinfo{journal}{\apj} \textbf{\bibinfo{volume}{592}}, \bibinfo{pages}{866}
  (\bibinfo{year}{2003}), \eprint{astro-ph/0304333}.

\bibitem[{\citenamefont{{Yoon} et~al.}(2006)\citenamefont{{Yoon}, {Yi}, and
  {Lee}}}]{yoon2006}
\bibinfo{author}{\bibfnamefont{S.-J.} \bibnamefont{{Yoon}}},
  \bibinfo{author}{\bibfnamefont{S.~K.} \bibnamefont{{Yi}}}, \bibnamefont{and}
  \bibinfo{author}{\bibfnamefont{Y.-W.} \bibnamefont{{Lee}}},
  \bibinfo{journal}{Science} \textbf{\bibinfo{volume}{311}},
  \bibinfo{pages}{1129} (\bibinfo{year}{2006}), \eprint{astro-ph/0601526}.

\bibitem[{\citenamefont{{Lee} et~al.}(2019)\citenamefont{{Lee}, {Chung}, and
  {Yoon}}}]{lee2019}
\bibinfo{author}{\bibfnamefont{S.-Y.} \bibnamefont{{Lee}}},
  \bibinfo{author}{\bibfnamefont{C.}~\bibnamefont{{Chung}}}, \bibnamefont{and}
  \bibinfo{author}{\bibfnamefont{S.-J.} \bibnamefont{{Yoon}}},
  \bibinfo{journal}{The Astrophysical Journal Supplement Series}
  \textbf{\bibinfo{volume}{240}}, \bibinfo{eid}{2} (\bibinfo{year}{2019}),
  \eprint{1811.00018}.

\bibitem[{\citenamefont{{Alves-Brito} et~al.}(2011)\citenamefont{{Alves-Brito},
  {Hau}, {Forbes}, {Spitler}, {Strader}, {Brodie}, and
  {Rhode}}}]{alves-brito2011}
\bibinfo{author}{\bibfnamefont{A.}~\bibnamefont{{Alves-Brito}}},
  \bibinfo{author}{\bibfnamefont{G.~K.~T.} \bibnamefont{{Hau}}},
  \bibinfo{author}{\bibfnamefont{D.~A.} \bibnamefont{{Forbes}}},
  \bibinfo{author}{\bibfnamefont{L.~R.} \bibnamefont{{Spitler}}},
  \bibinfo{author}{\bibfnamefont{J.}~\bibnamefont{{Strader}}},
  \bibinfo{author}{\bibfnamefont{J.~P.} \bibnamefont{{Brodie}}},
  \bibnamefont{and} \bibinfo{author}{\bibfnamefont{K.~L.}
  \bibnamefont{{Rhode}}}, \bibinfo{journal}{\mnras}
  \textbf{\bibinfo{volume}{417}}, \bibinfo{pages}{1823} (\bibinfo{year}{2011}),
  \eprint{1107.0757}.

\bibitem[{\citenamefont{{Usher} et~al.}(2012)\citenamefont{{Usher}, {Forbes},
  {Brodie}, {Foster}, {Spitler}, {Arnold}, {Romanowsky}, {Strader}, and
  {Pota}}}]{usher2012}
\bibinfo{author}{\bibfnamefont{C.}~\bibnamefont{{Usher}}},
  \bibinfo{author}{\bibfnamefont{D.~A.} \bibnamefont{{Forbes}}},
  \bibinfo{author}{\bibfnamefont{J.~P.} \bibnamefont{{Brodie}}},
  \bibinfo{author}{\bibfnamefont{C.}~\bibnamefont{{Foster}}},
  \bibinfo{author}{\bibfnamefont{L.~R.} \bibnamefont{{Spitler}}},
  \bibinfo{author}{\bibfnamefont{J.~A.} \bibnamefont{{Arnold}}},
  \bibinfo{author}{\bibfnamefont{A.~J.} \bibnamefont{{Romanowsky}}},
  \bibinfo{author}{\bibfnamefont{J.}~\bibnamefont{{Strader}}},
  \bibnamefont{and} \bibinfo{author}{\bibfnamefont{V.}~\bibnamefont{{Pota}}},
  \bibinfo{journal}{\mnras} \textbf{\bibinfo{volume}{426}},
  \bibinfo{pages}{1475} (\bibinfo{year}{2012}), \eprint{1207.6402}.

\bibitem[{\citenamefont{{Brodie} et~al.}(2012)\citenamefont{{Brodie}, {Usher},
  {Conroy}, {Strader}, {Arnold}, {Forbes}, and {Romanowsky}}}]{brodie2012}
\bibinfo{author}{\bibfnamefont{J.~P.} \bibnamefont{{Brodie}}},
  \bibinfo{author}{\bibfnamefont{C.}~\bibnamefont{{Usher}}},
  \bibinfo{author}{\bibfnamefont{C.}~\bibnamefont{{Conroy}}},
  \bibinfo{author}{\bibfnamefont{J.}~\bibnamefont{{Strader}}},
  \bibinfo{author}{\bibfnamefont{J.~A.} \bibnamefont{{Arnold}}},
  \bibinfo{author}{\bibfnamefont{D.~A.} \bibnamefont{{Forbes}}},
  \bibnamefont{and} \bibinfo{author}{\bibfnamefont{A.~J.}
  \bibnamefont{{Romanowsky}}}, \bibinfo{journal}{\apj}
  \textbf{\bibinfo{volume}{759}}, \bibinfo{eid}{L33} (\bibinfo{year}{2012}),
  \eprint{1209.5390}.

\bibitem[{\citenamefont{{Conroy} et~al.}(2014)\citenamefont{{Conroy}, {Graves},
  and {van Dokkum}}}]{conroy2014}
\bibinfo{author}{\bibfnamefont{C.}~\bibnamefont{{Conroy}}},
  \bibinfo{author}{\bibfnamefont{G.~J.} \bibnamefont{{Graves}}},
  \bibnamefont{and} \bibinfo{author}{\bibfnamefont{P.~G.} \bibnamefont{{van
  Dokkum}}}, \bibinfo{journal}{\apj} \textbf{\bibinfo{volume}{780}},
  \bibinfo{eid}{33} (\bibinfo{year}{2014}), \eprint{1303.6629}.

\bibitem[{\citenamefont{{Choi} et~al.}(2014)\citenamefont{{Choi}, {Conroy},
  {Moustakas}, {Graves}, {Holden}, {Brodwin}, {Brown}, and {van
  Dokkum}}}]{choi2014}
\bibinfo{author}{\bibfnamefont{J.}~\bibnamefont{{Choi}}},
  \bibinfo{author}{\bibfnamefont{C.}~\bibnamefont{{Conroy}}},
  \bibinfo{author}{\bibfnamefont{J.}~\bibnamefont{{Moustakas}}},
  \bibinfo{author}{\bibfnamefont{G.~J.} \bibnamefont{{Graves}}},
  \bibinfo{author}{\bibfnamefont{B.~P.} \bibnamefont{{Holden}}},
  \bibinfo{author}{\bibfnamefont{M.}~\bibnamefont{{Brodwin}}},
  \bibinfo{author}{\bibfnamefont{M.~J.~I.} \bibnamefont{{Brown}}},
  \bibnamefont{and} \bibinfo{author}{\bibfnamefont{P.~G.} \bibnamefont{{van
  Dokkum}}}, \bibinfo{journal}{\apj} \textbf{\bibinfo{volume}{792}},
  \bibinfo{eid}{95} (\bibinfo{year}{2014}), \eprint{1403.4932}.

\bibitem[{\citenamefont{{Conroy} et~al.}(2018)\citenamefont{{Conroy},
  {Villaume}, {van Dokkum}, and {Lind}}}]{conroy2018a}
\bibinfo{author}{\bibfnamefont{C.}~\bibnamefont{{Conroy}}},
  \bibinfo{author}{\bibfnamefont{A.}~\bibnamefont{{Villaume}}},
  \bibinfo{author}{\bibfnamefont{P.~G.} \bibnamefont{{van Dokkum}}},
  \bibnamefont{and} \bibinfo{author}{\bibfnamefont{K.}~\bibnamefont{{Lind}}},
  \bibinfo{journal}{\apj} \textbf{\bibinfo{volume}{854}}, \bibinfo{eid}{139}
  (\bibinfo{year}{2018}).

\bibitem[{\citenamefont{{Strader} et~al.}(2011)\citenamefont{{Strader},
  {Romanowsky}, {Brodie}, {Spitler}, {Beasley}, {Arnold}, {Tamura}, {Sharples},
  and {Arimoto}}}]{strader2011}
\bibinfo{author}{\bibfnamefont{J.}~\bibnamefont{{Strader}}},
  \bibinfo{author}{\bibfnamefont{A.~J.} \bibnamefont{{Romanowsky}}},
  \bibinfo{author}{\bibfnamefont{J.~P.} \bibnamefont{{Brodie}}},
  \bibinfo{author}{\bibfnamefont{L.~R.} \bibnamefont{{Spitler}}},
  \bibinfo{author}{\bibfnamefont{M.~A.} \bibnamefont{{Beasley}}},
  \bibinfo{author}{\bibfnamefont{J.~A.} \bibnamefont{{Arnold}}},
  \bibinfo{author}{\bibfnamefont{N.}~\bibnamefont{{Tamura}}},
  \bibinfo{author}{\bibfnamefont{R.~M.} \bibnamefont{{Sharples}}},
  \bibnamefont{and}
  \bibinfo{author}{\bibfnamefont{N.}~\bibnamefont{{Arimoto}}},
  \bibinfo{journal}{The Astrophysical Journal Supplement Series}
  \textbf{\bibinfo{volume}{197}}, \bibinfo{eid}{33} (\bibinfo{year}{2011}),
  \eprint{1110.2778}.

\bibitem[{\citenamefont{{Mihos} et~al.}(2017)\citenamefont{{Mihos}, {Harding},
  {Feldmeier}, {Rudick}, {Janowiecki}, {Morrison}, {Slater}, and
  {Watkins}}}]{mihos2017}
\bibinfo{author}{\bibfnamefont{J.~C.} \bibnamefont{{Mihos}}},
  \bibinfo{author}{\bibfnamefont{P.}~\bibnamefont{{Harding}}},
  \bibinfo{author}{\bibfnamefont{J.~J.} \bibnamefont{{Feldmeier}}},
  \bibinfo{author}{\bibfnamefont{C.}~\bibnamefont{{Rudick}}},
  \bibinfo{author}{\bibfnamefont{S.}~\bibnamefont{{Janowiecki}}},
  \bibinfo{author}{\bibfnamefont{H.}~\bibnamefont{{Morrison}}},
  \bibinfo{author}{\bibfnamefont{C.}~\bibnamefont{{Slater}}}, \bibnamefont{and}
  \bibinfo{author}{\bibfnamefont{A.}~\bibnamefont{{Watkins}}},
  \bibinfo{journal}{\apj} \textbf{\bibinfo{volume}{834}}, \bibinfo{eid}{16}
  (\bibinfo{year}{2017}), \eprint{1611.04435}.

\bibitem[{\citenamefont{{Oldham} and {Auger}}(2016)}]{oldham2016}
\bibinfo{author}{\bibfnamefont{L.~J.} \bibnamefont{{Oldham}}} \bibnamefont{and}
  \bibinfo{author}{\bibfnamefont{M.~W.} \bibnamefont{{Auger}}},
  \bibinfo{journal}{\mnras} \textbf{\bibinfo{volume}{455}},
  \bibinfo{pages}{820} (\bibinfo{year}{2016}).

\bibitem[{\citenamefont{{Jord{\'a}n} et~al.}(2009)\citenamefont{{Jord{\'a}n},
  {Peng}, {Blakeslee}, {C{\^o}t{\'e}}, {Eyheramendy}, {Ferrarese}, {Mei},
  {Tonry}, and {West}}}]{jordan2009}
\bibinfo{author}{\bibfnamefont{A.}~\bibnamefont{{Jord{\'a}n}}},
  \bibinfo{author}{\bibfnamefont{E.~W.} \bibnamefont{{Peng}}},
  \bibinfo{author}{\bibfnamefont{J.~P.} \bibnamefont{{Blakeslee}}},
  \bibinfo{author}{\bibfnamefont{P.}~\bibnamefont{{C{\^o}t{\'e}}}},
  \bibinfo{author}{\bibfnamefont{S.}~\bibnamefont{{Eyheramendy}}},
  \bibinfo{author}{\bibfnamefont{L.}~\bibnamefont{{Ferrarese}}},
  \bibinfo{author}{\bibfnamefont{S.}~\bibnamefont{{Mei}}},
  \bibinfo{author}{\bibfnamefont{J.~L.} \bibnamefont{{Tonry}}},
  \bibnamefont{and} \bibinfo{author}{\bibfnamefont{M.~J.}
  \bibnamefont{{West}}}, \bibinfo{journal}{The Astrophysical Journal Supplement
  Series} \textbf{\bibinfo{volume}{180}}, \bibinfo{pages}{54}
  (\bibinfo{year}{2009}).

\bibitem[{\citenamefont{{Villaume}
  et~al.}(2017{\natexlab{a}})\citenamefont{{Villaume}, {Conroy}, {Johnson},
  {Rayner}, {Mann}, and {van Dokkum}}}]{villaume2017a}
\bibinfo{author}{\bibfnamefont{A.}~\bibnamefont{{Villaume}}},
  \bibinfo{author}{\bibfnamefont{C.}~\bibnamefont{{Conroy}}},
  \bibinfo{author}{\bibfnamefont{B.}~\bibnamefont{{Johnson}}},
  \bibinfo{author}{\bibfnamefont{J.}~\bibnamefont{{Rayner}}},
  \bibinfo{author}{\bibfnamefont{A.~W.} \bibnamefont{{Mann}}},
  \bibnamefont{and} \bibinfo{author}{\bibfnamefont{P.}~\bibnamefont{{van
  Dokkum}}}, \bibinfo{journal}{The Astrophysical Journal Supplement Series}
  \textbf{\bibinfo{volume}{230}}, \bibinfo{eid}{23}
  (\bibinfo{year}{2017}{\natexlab{a}}).

\bibitem[{\citenamefont{{S{\'a}nchez-Bl{\'a}zquez}
  et~al.}(2006)\citenamefont{{S{\'a}nchez-Bl{\'a}zquez}, {Peletier},
  {Jim{\'e}nez- Vicente}, {Cardiel}, {Cenarro}, {Falc{\'o}n-Barroso}, {Gorgas},
  {Selam}, and {Vazdekis}}}]{sb2006a}
\bibinfo{author}{\bibfnamefont{P.}~\bibnamefont{{S{\'a}nchez-Bl{\'a}zquez}}},
  \bibinfo{author}{\bibfnamefont{R.~F.} \bibnamefont{{Peletier}}},
  \bibinfo{author}{\bibfnamefont{J.}~\bibnamefont{{Jim{\'e}nez- Vicente}}},
  \bibinfo{author}{\bibfnamefont{N.}~\bibnamefont{{Cardiel}}},
  \bibinfo{author}{\bibfnamefont{A.~J.} \bibnamefont{{Cenarro}}},
  \bibinfo{author}{\bibfnamefont{J.}~\bibnamefont{{Falc{\'o}n-Barroso}}},
  \bibinfo{author}{\bibfnamefont{J.}~\bibnamefont{{Gorgas}}},
  \bibinfo{author}{\bibfnamefont{S.}~\bibnamefont{{Selam}}}, \bibnamefont{and}
  \bibinfo{author}{\bibfnamefont{A.}~\bibnamefont{{Vazdekis}}},
  \bibinfo{journal}{\mnras} \textbf{\bibinfo{volume}{371}},
  \bibinfo{pages}{703} (\bibinfo{year}{2006}), \eprint{astro-ph/0607009}.

\bibitem[{\citenamefont{{Mann} et~al.}(2015)\citenamefont{{Mann}, {Feiden},
  {Gaidos}, {Boyajian}, and {von Braun}}}]{mann2015}
\bibinfo{author}{\bibfnamefont{A.~W.} \bibnamefont{{Mann}}},
  \bibinfo{author}{\bibfnamefont{G.~A.} \bibnamefont{{Feiden}}},
  \bibinfo{author}{\bibfnamefont{E.}~\bibnamefont{{Gaidos}}},
  \bibinfo{author}{\bibfnamefont{T.}~\bibnamefont{{Boyajian}}},
  \bibnamefont{and} \bibinfo{author}{\bibfnamefont{K.}~\bibnamefont{{von
  Braun}}}, \bibinfo{journal}{\apj} \textbf{\bibinfo{volume}{804}},
  \bibinfo{eid}{64} (\bibinfo{year}{2015}), \eprint{1501.01635}.

\bibitem[{\citenamefont{{Schiavon} et~al.}(2005)\citenamefont{{Schiavon},
  {Rose}, {Courteau}, and {MacArthur}}}]{schiavon2005}
\bibinfo{author}{\bibfnamefont{R.~P.} \bibnamefont{{Schiavon}}},
  \bibinfo{author}{\bibfnamefont{J.~A.} \bibnamefont{{Rose}}},
  \bibinfo{author}{\bibfnamefont{S.}~\bibnamefont{{Courteau}}},
  \bibnamefont{and} \bibinfo{author}{\bibfnamefont{L.~A.}
  \bibnamefont{{MacArthur}}}, \bibinfo{journal}{The Astrophysical Journal
  Supplement Series} \textbf{\bibinfo{volume}{160}}, \bibinfo{pages}{163}
  (\bibinfo{year}{2005}), \eprint{astro-ph/0504313}.

\bibitem[{\citenamefont{{Roediger} et~al.}(2014)\citenamefont{{Roediger},
  {Courteau}, {Graves}, and {Schiavon}}}]{roediger2014}
\bibinfo{author}{\bibfnamefont{J.~C.} \bibnamefont{{Roediger}}},
  \bibinfo{author}{\bibfnamefont{S.}~\bibnamefont{{Courteau}}},
  \bibinfo{author}{\bibfnamefont{G.}~\bibnamefont{{Graves}}}, \bibnamefont{and}
  \bibinfo{author}{\bibfnamefont{R.~P.} \bibnamefont{{Schiavon}}},
  \bibinfo{journal}{The Astrophysical Journal Supplement Series}
  \textbf{\bibinfo{volume}{210}}, \bibinfo{eid}{10} (\bibinfo{year}{2014}),
  \eprint{1310.3275}.

\bibitem[{\citenamefont{{S{\'a}nchez-Bl{\'a}zquez}
  et~al.}(2011)\citenamefont{{S{\'a}nchez-Bl{\'a}zquez}, {Ocvirk}, {Gibson},
  {P{\'e}rez}, and {Peletier}}}]{sb2011a}
\bibinfo{author}{\bibfnamefont{P.}~\bibnamefont{{S{\'a}nchez-Bl{\'a}zquez}}},
  \bibinfo{author}{\bibfnamefont{P.}~\bibnamefont{{Ocvirk}}},
  \bibinfo{author}{\bibfnamefont{B.~K.} \bibnamefont{{Gibson}}},
  \bibinfo{author}{\bibfnamefont{I.}~\bibnamefont{{P{\'e}rez}}},
  \bibnamefont{and} \bibinfo{author}{\bibfnamefont{R.~F.}
  \bibnamefont{{Peletier}}}, \bibinfo{journal}{\mnras}
  \textbf{\bibinfo{volume}{415}}, \bibinfo{pages}{709} (\bibinfo{year}{2011}),
  \eprint{1103.3796}.

\bibitem[{\citenamefont{{Faber} et~al.}(1985)\citenamefont{{Faber}, {Friel},
  {Burstein}, and {Gaskell}}}]{faber1985}
\bibinfo{author}{\bibfnamefont{S.~M.} \bibnamefont{{Faber}}},
  \bibinfo{author}{\bibfnamefont{E.~D.} \bibnamefont{{Friel}}},
  \bibinfo{author}{\bibfnamefont{D.}~\bibnamefont{{Burstein}}},
  \bibnamefont{and} \bibinfo{author}{\bibfnamefont{C.~M.}
  \bibnamefont{{Gaskell}}}, \bibinfo{journal}{The Astrophysical Journal
  Supplement Series} \textbf{\bibinfo{volume}{57}}, \bibinfo{pages}{711}
  (\bibinfo{year}{1985}).

\bibitem[{\citenamefont{{Burstein} et~al.}(1986)\citenamefont{{Burstein},
  {Faber}, and {Gonzalez}}}]{burstein1986}
\bibinfo{author}{\bibfnamefont{D.}~\bibnamefont{{Burstein}}},
  \bibinfo{author}{\bibfnamefont{S.~M.} \bibnamefont{{Faber}}},
  \bibnamefont{and} \bibinfo{author}{\bibfnamefont{J.~J.}
  \bibnamefont{{Gonzalez}}}, \bibinfo{journal}{\aj}
  \textbf{\bibinfo{volume}{91}}, \bibinfo{pages}{1130} (\bibinfo{year}{1986}).

\bibitem[{\citenamefont{{Worthey} et~al.}(1994)\citenamefont{{Worthey},
  {Faber}, {Gonzalez}, and {Burstein}}}]{worthey1994}
\bibinfo{author}{\bibfnamefont{G.}~\bibnamefont{{Worthey}}},
  \bibinfo{author}{\bibfnamefont{S.~M.} \bibnamefont{{Faber}}},
  \bibinfo{author}{\bibfnamefont{J.~J.} \bibnamefont{{Gonzalez}}},
  \bibnamefont{and}
  \bibinfo{author}{\bibfnamefont{D.}~\bibnamefont{{Burstein}}},
  \bibinfo{journal}{The Astrophysical Journal Supplement Series}
  \textbf{\bibinfo{volume}{94}}, \bibinfo{pages}{687} (\bibinfo{year}{1994}).

\bibitem[{\citenamefont{{Cohen} and {Ryzhov}}(1997)}]{cohen1997}
\bibinfo{author}{\bibfnamefont{J.~G.} \bibnamefont{{Cohen}}} \bibnamefont{and}
  \bibinfo{author}{\bibfnamefont{A.}~\bibnamefont{{Ryzhov}}},
  \bibinfo{journal}{\apj} \textbf{\bibinfo{volume}{486}}, \bibinfo{pages}{230}
  (\bibinfo{year}{1997}), \eprint{astro-ph/9704051}.

\bibitem[{\citenamefont{{Hanes} et~al.}(2001)\citenamefont{{Hanes},
  {C{\^o}t{\'e}}, {Bridges}, {McLaughlin}, {Geisler}, {Harris}, {Hesser}, and
  {Lee}}}]{hanes2001}
\bibinfo{author}{\bibfnamefont{D.~A.} \bibnamefont{{Hanes}}},
  \bibinfo{author}{\bibfnamefont{P.}~\bibnamefont{{C{\^o}t{\'e}}}},
  \bibinfo{author}{\bibfnamefont{T.~J.} \bibnamefont{{Bridges}}},
  \bibinfo{author}{\bibfnamefont{D.~E.} \bibnamefont{{McLaughlin}}},
  \bibinfo{author}{\bibfnamefont{D.}~\bibnamefont{{Geisler}}},
  \bibinfo{author}{\bibfnamefont{G.~L.~H.} \bibnamefont{{Harris}}},
  \bibinfo{author}{\bibfnamefont{J.~E.} \bibnamefont{{Hesser}}},
  \bibnamefont{and} \bibinfo{author}{\bibfnamefont{M.~G.} \bibnamefont{{Lee}}},
  \bibinfo{journal}{\apj} \textbf{\bibinfo{volume}{559}}, \bibinfo{pages}{812}
  (\bibinfo{year}{2001}), \eprint{astro-ph/0106004}.

\bibitem[{\citenamefont{{Strom} et~al.}(1981)\citenamefont{{Strom}, {Forte},
  {Harris}, {Strom}, {Wells}, and {Smith}}}]{strom1981}
\bibinfo{author}{\bibfnamefont{S.~E.} \bibnamefont{{Strom}}},
  \bibinfo{author}{\bibfnamefont{J.~C.} \bibnamefont{{Forte}}},
  \bibinfo{author}{\bibfnamefont{W.~E.} \bibnamefont{{Harris}}},
  \bibinfo{author}{\bibfnamefont{K.~M.} \bibnamefont{{Strom}}},
  \bibinfo{author}{\bibfnamefont{D.~C.} \bibnamefont{{Wells}}},
  \bibnamefont{and} \bibinfo{author}{\bibfnamefont{M.~G.}
  \bibnamefont{{Smith}}}, \bibinfo{journal}{\apj}
  \textbf{\bibinfo{volume}{245}}, \bibinfo{pages}{416} (\bibinfo{year}{1981}).

\bibitem[{\citenamefont{{Fitzpatrick}}(1999)}]{fitzpatrick1999}
\bibinfo{author}{\bibfnamefont{E.~L.} \bibnamefont{{Fitzpatrick}}},
  \bibinfo{journal}{Publications of the Astronomical Society of the Pacific}
  \textbf{\bibinfo{volume}{111}}, \bibinfo{pages}{63} (\bibinfo{year}{1999}),
  \eprint{astro-ph/9809387}.

\bibitem[{\citenamefont{{Schlegel} et~al.}(1998)\citenamefont{{Schlegel},
  {Finkbeiner}, and {Davis}}}]{schlegel1998}
\bibinfo{author}{\bibfnamefont{D.~J.} \bibnamefont{{Schlegel}}},
  \bibinfo{author}{\bibfnamefont{D.~P.} \bibnamefont{{Finkbeiner}}},
  \bibnamefont{and} \bibinfo{author}{\bibfnamefont{M.}~\bibnamefont{{Davis}}},
  \bibinfo{journal}{\apj} \textbf{\bibinfo{volume}{500}}, \bibinfo{pages}{525}
  (\bibinfo{year}{1998}), \eprint{astro-ph/9710327}.

\bibitem[{\citenamefont{{Ferrarese} et~al.}(2012)\citenamefont{{Ferrarese},
  {C{\^o}t{\'e}}, {Cuilland re}, {Gwyn}, {Peng}, {MacArthur}, {Duc}, {Boselli},
  {Mei}, {Erben} et~al.}}]{ferrarese2012}
\bibinfo{author}{\bibfnamefont{L.}~\bibnamefont{{Ferrarese}}},
  \bibinfo{author}{\bibfnamefont{P.}~\bibnamefont{{C{\^o}t{\'e}}}},
  \bibinfo{author}{\bibfnamefont{J.-C.} \bibnamefont{{Cuilland re}}},
  \bibinfo{author}{\bibfnamefont{S.~D.~J.} \bibnamefont{{Gwyn}}},
  \bibinfo{author}{\bibfnamefont{E.~W.} \bibnamefont{{Peng}}},
  \bibinfo{author}{\bibfnamefont{L.~A.} \bibnamefont{{MacArthur}}},
  \bibinfo{author}{\bibfnamefont{P.-A.} \bibnamefont{{Duc}}},
  \bibinfo{author}{\bibfnamefont{A.}~\bibnamefont{{Boselli}}},
  \bibinfo{author}{\bibfnamefont{S.}~\bibnamefont{{Mei}}},
  \bibinfo{author}{\bibfnamefont{T.}~\bibnamefont{{Erben}}},
  \bibnamefont{et~al.}, \bibinfo{journal}{The Astrophysical Journal Supplement
  Series} \textbf{\bibinfo{volume}{200}}, \bibinfo{eid}{4}
  (\bibinfo{year}{2012}).

\bibitem[{\citenamefont{{Hogg} et~al.}(2010)\citenamefont{{Hogg}, {Bovy}, and
  {Lang}}}]{hogg2010}
\bibinfo{author}{\bibfnamefont{D.~W.} \bibnamefont{{Hogg}}},
  \bibinfo{author}{\bibfnamefont{J.}~\bibnamefont{{Bovy}}}, \bibnamefont{and}
  \bibinfo{author}{\bibfnamefont{D.}~\bibnamefont{{Lang}}},
  \bibinfo{journal}{ArXiv e-prints} \bibinfo{eid}{arXiv:1008.4686}
  (\bibinfo{year}{2010}), \eprint{1008.4686}.

\bibitem[{\citenamefont{{Harris}}(1996)}]{harris1996}
\bibinfo{author}{\bibfnamefont{W.~E.} \bibnamefont{{Harris}}},
  \bibinfo{journal}{\aj} \textbf{\bibinfo{volume}{112}}, \bibinfo{pages}{1487}
  (\bibinfo{year}{1996}).

\bibitem[{\citenamefont{{Zinn}}(1985)}]{zinn1985}
\bibinfo{author}{\bibfnamefont{R.}~\bibnamefont{{Zinn}}},
  \bibinfo{journal}{\apj} \textbf{\bibinfo{volume}{293}}, \bibinfo{pages}{424}
  (\bibinfo{year}{1985}).

\bibitem[{\citenamefont{{Choi} et~al.}(2016)\citenamefont{{Choi}, {Dotter},
  {Conroy}, {Cantiello}, {Paxton}, and {Johnson}}}]{choi2016}
\bibinfo{author}{\bibfnamefont{J.}~\bibnamefont{{Choi}}},
  \bibinfo{author}{\bibfnamefont{A.}~\bibnamefont{{Dotter}}},
  \bibinfo{author}{\bibfnamefont{C.}~\bibnamefont{{Conroy}}},
  \bibinfo{author}{\bibfnamefont{M.}~\bibnamefont{{Cantiello}}},
  \bibinfo{author}{\bibfnamefont{B.}~\bibnamefont{{Paxton}}}, \bibnamefont{and}
  \bibinfo{author}{\bibfnamefont{B.~D.} \bibnamefont{{Johnson}}},
  \bibinfo{journal}{\apj} \textbf{\bibinfo{volume}{823}}, \bibinfo{eid}{102}
  (\bibinfo{year}{2016}), \eprint{1604.08592}.

\bibitem[{\citenamefont{{Tripicco} and {Bell}}(1995)}]{tripicco1995}
\bibinfo{author}{\bibfnamefont{M.~J.} \bibnamefont{{Tripicco}}}
  \bibnamefont{and} \bibinfo{author}{\bibfnamefont{R.~A.}
  \bibnamefont{{Bell}}}, \bibinfo{journal}{\aj} \textbf{\bibinfo{volume}{110}},
  \bibinfo{pages}{3035} (\bibinfo{year}{1995}).

\bibitem[{\citenamefont{{Agnello} et~al.}(2014)\citenamefont{{Agnello},
  {Evans}, {Romanowsky}, and {Brodie}}}]{agnello2014b}
\bibinfo{author}{\bibfnamefont{A.}~\bibnamefont{{Agnello}}},
  \bibinfo{author}{\bibfnamefont{N.~W.} \bibnamefont{{Evans}}},
  \bibinfo{author}{\bibfnamefont{A.~J.} \bibnamefont{{Romanowsky}}},
  \bibnamefont{and} \bibinfo{author}{\bibfnamefont{J.~P.}
  \bibnamefont{{Brodie}}}, \bibinfo{journal}{\mnras}
  \textbf{\bibinfo{volume}{442}}, \bibinfo{pages}{3299} (\bibinfo{year}{2014}),
  \eprint{1401.4461}.

\bibitem[{\citenamefont{{Villaume}
  et~al.}(2017{\natexlab{b}})\citenamefont{{Villaume}, {Brodie}, {Conroy},
  {Romanowsky}, and {van Dokkum}}}]{villaume2017b}
\bibinfo{author}{\bibfnamefont{A.}~\bibnamefont{{Villaume}}},
  \bibinfo{author}{\bibfnamefont{J.}~\bibnamefont{{Brodie}}},
  \bibinfo{author}{\bibfnamefont{C.}~\bibnamefont{{Conroy}}},
  \bibinfo{author}{\bibfnamefont{A.~J.} \bibnamefont{{Romanowsky}}},
  \bibnamefont{and} \bibinfo{author}{\bibfnamefont{P.}~\bibnamefont{{van
  Dokkum}}}, \bibinfo{journal}{\apj} \textbf{\bibinfo{volume}{850}},
  \bibinfo{eid}{L14} (\bibinfo{year}{2017}{\natexlab{b}}).

\bibitem[{\citenamefont{{Romanowsky} et~al.}(2012)\citenamefont{{Romanowsky},
  {Strader}, {Brodie}, {Mihos}, {Spitler}, {Forbes}, {Foster}, and
  {Arnold}}}]{romanowsky2012}
\bibinfo{author}{\bibfnamefont{A.~J.} \bibnamefont{{Romanowsky}}},
  \bibinfo{author}{\bibfnamefont{J.}~\bibnamefont{{Strader}}},
  \bibinfo{author}{\bibfnamefont{J.~P.} \bibnamefont{{Brodie}}},
  \bibinfo{author}{\bibfnamefont{J.~C.} \bibnamefont{{Mihos}}},
  \bibinfo{author}{\bibfnamefont{L.~R.} \bibnamefont{{Spitler}}},
  \bibinfo{author}{\bibfnamefont{D.~A.} \bibnamefont{{Forbes}}},
  \bibinfo{author}{\bibfnamefont{C.}~\bibnamefont{{Foster}}}, \bibnamefont{and}
  \bibinfo{author}{\bibfnamefont{J.~A.} \bibnamefont{{Arnold}}},
  \bibinfo{journal}{\apj} \textbf{\bibinfo{volume}{748}}, \bibinfo{eid}{29}
  (\bibinfo{year}{2012}), \eprint{1112.3959}.

\bibitem[{\citenamefont{{Pfeffer} et~al.}(2018)\citenamefont{{Pfeffer},
  {Kruijssen}, {Crain}, and {Bastian}}}]{pfeffer2018}
\bibinfo{author}{\bibfnamefont{J.}~\bibnamefont{{Pfeffer}}},
  \bibinfo{author}{\bibfnamefont{J.~M.~D.} \bibnamefont{{Kruijssen}}},
  \bibinfo{author}{\bibfnamefont{R.~A.} \bibnamefont{{Crain}}},
  \bibnamefont{and}
  \bibinfo{author}{\bibfnamefont{N.}~\bibnamefont{{Bastian}}},
  \bibinfo{journal}{\mnras} \textbf{\bibinfo{volume}{475}},
  \bibinfo{pages}{4309} (\bibinfo{year}{2018}).

\bibitem[{\citenamefont{{Choksi} et~al.}(2018)\citenamefont{{Choksi}, {Gnedin},
  and {Li}}}]{choksi2018}
\bibinfo{author}{\bibfnamefont{N.}~\bibnamefont{{Choksi}}},
  \bibinfo{author}{\bibfnamefont{O.~Y.} \bibnamefont{{Gnedin}}},
  \bibnamefont{and} \bibinfo{author}{\bibfnamefont{H.}~\bibnamefont{{Li}}},
  \bibinfo{journal}{\mnras} \textbf{\bibinfo{volume}{480}},
  \bibinfo{pages}{2343} (\bibinfo{year}{2018}), \eprint{1801.03515}.

\bibitem[{\citenamefont{{Strader} et~al.}(2004)\citenamefont{{Strader},
  {Brodie}, and {Forbes}}}]{strader2004}
\bibinfo{author}{\bibfnamefont{J.}~\bibnamefont{{Strader}}},
  \bibinfo{author}{\bibfnamefont{J.~P.} \bibnamefont{{Brodie}}},
  \bibnamefont{and} \bibinfo{author}{\bibfnamefont{D.~A.}
  \bibnamefont{{Forbes}}}, \bibinfo{journal}{\aj}
  \textbf{\bibinfo{volume}{127}}, \bibinfo{pages}{3431} (\bibinfo{year}{2004}),
  \eprint{astro-ph/0403160}.

\bibitem[{\citenamefont{{Kundu} and
  {Whitmore}}(2001{\natexlab{b}})}]{kundu2001b}
\bibinfo{author}{\bibfnamefont{A.}~\bibnamefont{{Kundu}}} \bibnamefont{and}
  \bibinfo{author}{\bibfnamefont{B.~C.} \bibnamefont{{Whitmore}}},
  \bibinfo{journal}{\aj} \textbf{\bibinfo{volume}{122}}, \bibinfo{pages}{1251}
  (\bibinfo{year}{2001}{\natexlab{b}}), \eprint{astro-ph/0105198}.

\bibitem[{\citenamefont{{Barmby} et~al.}(2000)\citenamefont{{Barmby}, {Huchra},
  {Brodie}, {Forbes}, {Schroder}, and {Grillmair}}}]{barmby2000}
\bibinfo{author}{\bibfnamefont{P.}~\bibnamefont{{Barmby}}},
  \bibinfo{author}{\bibfnamefont{J.~P.} \bibnamefont{{Huchra}}},
  \bibinfo{author}{\bibfnamefont{J.~P.} \bibnamefont{{Brodie}}},
  \bibinfo{author}{\bibfnamefont{D.~A.} \bibnamefont{{Forbes}}},
  \bibinfo{author}{\bibfnamefont{L.~L.} \bibnamefont{{Schroder}}},
  \bibnamefont{and} \bibinfo{author}{\bibfnamefont{C.~J.}
  \bibnamefont{{Grillmair}}}, \bibinfo{journal}{\aj}
  \textbf{\bibinfo{volume}{119}}, \bibinfo{pages}{727} (\bibinfo{year}{2000}),
  \eprint{astro-ph/9911152}.

\bibitem[{\citenamefont{{Robertson} et~al.}(2005)\citenamefont{{Robertson},
  {Bullock}, {Font}, {Johnston}, and {Hernquist}}}]{robertson2005}
\bibinfo{author}{\bibfnamefont{B.}~\bibnamefont{{Robertson}}},
  \bibinfo{author}{\bibfnamefont{J.~S.} \bibnamefont{{Bullock}}},
  \bibinfo{author}{\bibfnamefont{A.~S.} \bibnamefont{{Font}}},
  \bibinfo{author}{\bibfnamefont{K.~V.} \bibnamefont{{Johnston}}},
  \bibnamefont{and}
  \bibinfo{author}{\bibfnamefont{L.}~\bibnamefont{{Hernquist}}},
  \bibinfo{journal}{\apj} \textbf{\bibinfo{volume}{632}}, \bibinfo{pages}{872}
  (\bibinfo{year}{2005}), \eprint{astro-ph/0501398}.

\bibitem[{\citenamefont{{Powalka} et~al.}(2016)\citenamefont{{Powalka},
  {Puzia}, {Lan{\c{c}}on}, {Peng}, {Sch{\"o}nebeck}, {Alamo-Mart{\'\i}nez},
  {{\'A}ngel}, {Blakeslee}, {C{\^o}t{\'e}}, {Cuilland re}
  et~al.}}]{powalka2016a}
\bibinfo{author}{\bibfnamefont{M.}~\bibnamefont{{Powalka}}},
  \bibinfo{author}{\bibfnamefont{T.~H.} \bibnamefont{{Puzia}}},
  \bibinfo{author}{\bibfnamefont{A.}~\bibnamefont{{Lan{\c{c}}on}}},
  \bibinfo{author}{\bibfnamefont{E.~W.} \bibnamefont{{Peng}}},
  \bibinfo{author}{\bibfnamefont{F.}~\bibnamefont{{Sch{\"o}nebeck}}},
  \bibinfo{author}{\bibfnamefont{K.}~\bibnamefont{{Alamo-Mart{\'\i}nez}}},
  \bibinfo{author}{\bibfnamefont{S.}~\bibnamefont{{{\'A}ngel}}},
  \bibinfo{author}{\bibfnamefont{J.~P.} \bibnamefont{{Blakeslee}}},
  \bibinfo{author}{\bibfnamefont{P.}~\bibnamefont{{C{\^o}t{\'e}}}},
  \bibinfo{author}{\bibfnamefont{J.-C.} \bibnamefont{{Cuilland re}}},
  \bibnamefont{et~al.}, \bibinfo{journal}{\apj} \textbf{\bibinfo{volume}{829}},
  \bibinfo{eid}{L5} (\bibinfo{year}{2016}), \eprint{1608.08628}.

\bibitem[{\citenamefont{P\'erez and Granger}(2007)}]{PER-GRA:2007}
\bibinfo{author}{\bibfnamefont{F.}~\bibnamefont{P\'erez}} \bibnamefont{and}
  \bibinfo{author}{\bibfnamefont{B.~E.} \bibnamefont{Granger}},
  \bibinfo{journal}{Computing in Science and Engineering}
  \textbf{\bibinfo{volume}{9}}, \bibinfo{pages}{21} (\bibinfo{year}{2007}),
  ISSN \bibinfo{issn}{1521-9615}, \urlprefix\url{http://ipython.org}.

\bibitem[{jon(2001)}]{jones_scipy_2001}
\emph{\bibinfo{title}{{SciPy}: Open source scientific tools for python}}
  (\bibinfo{year}{2001}), \urlprefix\url{http://www.scipy.org/}.

\bibitem[{\citenamefont{Van Der~Walt et~al.}(2011)\citenamefont{Van Der~Walt,
  Colbert, and Varoquaux}}]{van2011numpy}
\bibinfo{author}{\bibfnamefont{S.}~\bibnamefont{Van Der~Walt}},
  \bibinfo{author}{\bibfnamefont{S.~C.} \bibnamefont{Colbert}},
  \bibnamefont{and}
  \bibinfo{author}{\bibfnamefont{G.}~\bibnamefont{Varoquaux}},
  \bibinfo{journal}{Computing in Science \& Engineering}
  \textbf{\bibinfo{volume}{13}}, \bibinfo{pages}{22} (\bibinfo{year}{2011}).

\bibitem[{\citenamefont{Hunter}(2007)}]{Hunter:2007}
\bibinfo{author}{\bibfnamefont{J.~D.} \bibnamefont{Hunter}},
  \bibinfo{journal}{Computing In Science \& Engineering}
  \textbf{\bibinfo{volume}{9}}, \bibinfo{pages}{90} (\bibinfo{year}{2007}).

\bibitem[{\citenamefont{{Taylor}}(2005)}]{topcat}
\bibinfo{author}{\bibfnamefont{M.~B.} \bibnamefont{{Taylor}}}, in
  \emph{\bibinfo{booktitle}{Astronomical Data Analysis Software and Systems
  XIV}}, edited by
  \bibinfo{editor}{\bibfnamefont{P.}~\bibnamefont{{Shopbell}}},
  \bibinfo{editor}{\bibfnamefont{M.}~\bibnamefont{{Britton}}},
  \bibnamefont{and} \bibinfo{editor}{\bibfnamefont{R.}~\bibnamefont{{Ebert}}}
  (\bibinfo{year}{2005}), vol. \bibinfo{volume}{347} of
  \emph{\bibinfo{series}{Astronomical Society of the Pacific Conference
  Series}}, p.~\bibinfo{pages}{29}.

\end{thebibliography}

\acknowledgements
The authors would like to thank A. L. Chies Santos, C. Conroy, S. Faber, and A. Wasserman for helpful discussions. AV was supported by an NSF Graduate Research Fellowship. JB was supported by  National Science Foundation grants AST-1616598 and AST-1518294. AJR was supported by National Science Foundation grants AST-1515084 and AST-1616710, and as a Research Corporation for Science Advancement Cottrell Scholar. JS was supported by NSF grant AST-1514763 and the Packard Foundation. This research has made use of the NASA/ IPAC Infrared Science Archive, which is operated by the Jet Propulsion Laboratory, California Institute of Technology, under contract with the National Aeronautics and Space Administration. This work made use of the IPython package \citep{PER-GRA:2007}, SciPy \citep{jones_scipy_2001}, NumPy \citep{van2011numpy}, matplotlib, a Python library for publication quality graphics \citep{Hunter:2007}, and of TOPCAT \citep{topcat}.

\end{document}